\documentclass[psfig]{mn2e}
\usepackage{graphicx}
\title[Colour Gradients of BCGs and E/S0 Galaxies]
 {Colour Gradients and the Colour-Magnitude Relation: Different Properties of Brightest Cluster Galaxies and E/S0 Galaxies in the Sloan Digital Sky Survey} 
\author[N. Roche, M. Bernardi and J. Hyde] 
 {Nathan Roche$^1$,Mariangela Bernardi and Joseph Hyde 
 \thanks{nathanroche@mac.com; bernardi@sas.upenn.edu; jhyde@sas.upenn.edu}\\
  University of Pennsylvania, 209 South 33rd Street, 
  Physics and Astronomy, Philadelphia, PA 19104, USA.\\
 $^1$ now at Osservatorio Astronomico di Bologna, Via Ranzani 1. Bologna 40127, 
Italy.}

\begin{document}

\date{}

\pagerange{\pageref{firstpage}--\pageref{lastpage}} \pubyear{2009}

\maketitle

\label{firstpage}

\begin{abstract} 
We examine the colour-magnitude relation of $\sim 5000$ Brightest Cluster Galaxies (BCGs) in the SDSS. The colour-magnitude and colour-velocity dispersion relations of the BCGs are flatter in slope, by factors of two or more, than those of non-BCG early-type galaxies of similar luminosity ($M_r<-22.5$), while their $g-r$ colours at the half-light radius are $\sim 0.01$ magnitude redder. 

 We investigate and compare radial colour gradients (which are usually negative) in these BCGs and a sample of 37000 early-type galaxies, using a gradient estimator based on the ratio of de Vaucouleurs effective radii in $g$ and $r$ passbands, ${{{\rm r}_{eff}(g)}\over{{\rm r}_{eff}(r)}}-1$. The mean colour gradient of BCGs is flatter (by $23\%$) than for other E/S0 galaxies of similar luminosity.

 The colour gradients in early-type galaxies are stronger at intermediate luminosity ($M_r\simeq -22$) than at low or the highest luminosities, and tend to decrease with increasing velocity dispersion ($\sigma$). In non-BCG E/S0s, colour gradients increase for larger effective radii (up to 10--12 kpc), and are negatively correlated with $10~{\rm log}~\sigma+M_r$, the mass density, and stellar age. However, gradients can be reduced or inverted (positive) for post-starburst galaxies at the youngest ages.
In BCGs, these trends are absent and the mean colour gradient remains at a relatively low level ($\sim 0.08$ by our measure) whatever the other properties of the galaxies. The redder half-light radius colours of the BCGs can be explained by their slightly greater ages combined with flatter radial colour gradients.

 We discuss possible explanations in terms of spheroidal galaxies initially having a colour gradient positively correlated with luminosity and positively correlated with large radius and/or low density. Subsequently, elliptical-elliptical `dry' mergers progressively reduce colour gradients, towards a low but non-zero value. This has occurred a greater number of times during the formation histories of the most massive E/S0s, and to by far the greatest degree in the BCGs.   
\end{abstract}


\section{Introduction} 
In this paper we investigate colour-magnitude relations, and, to greater depth, the properties of radial colour gradients, for early-type galaxies and in particular Brightest Cluster Galaxies (BCGs). Both can provide information on these objects' formation.

Early-type galaxies (elliptical and S0) with higher luminosities are redder in colour than those of lower luminosity, following a relatively tight and linear  colour-magnitude relation (CMR), or `red sequence'. Gallazzi et al. (2005) performed a detailed analysis of $2\times 10^5$ galaxy spectra from the Sloan Digital Sky Survey (SDSS), using a set of 5 spectral indices (D4000, $\rm H\beta$, $\rm H_{\delta A}+H_{\gamma A}$, $[\rm Mg_2 Fe]$ and $\rm [MgFe]^{\prime}$) to estimate mean stellar ages and metallicities. Gallazzi et al. (2006) and Jimenez et al. (2007), from spectral analysis of the early-type galaxies, conclude that the slope of the red-sequence CMR is primarily caused by an increase in mean stellar metallicity with galaxy mass or velocity dispersion. Mergers may then affect the CMR by building  more luminous galaxies with the same composition as their less massive progenitors, which would tend to flatten the CMR slope at $L>L^*$ luminosities (eg. Skelton, Bell and Somerville 2009). 

 Age also has a (secondary) role in explaining the colour trend - the most massive E/S0s formed early and rapidly (in $<1$ Gyr), while those of lower luminosity formed stars over longer timescales, giving slightly younger flux-weighted ages. There is also evidence that E/S0 galaxy formation occurred slightly earlier in dense environments (eg. Sheth et al. 2006, Bernardi 2009), while galaxies in cluster centres may have had an unusual and extreme merger history (e.g. Boylan-Kolchin, Ma and Quataert 2006).  Thus merger history and cluster environment may both influence the CMR.

Our studies of the CMR in Roche, Bernardi and Hyde 2009 (hereafter Paper I)
suggested (see Section 3.1) that the radial colour gradients in E/S0 galaxies might be correlated with their other properties and it would be fruitful to investigate this further. These colour gradients are of interest in that they are sensitive tracers of evolution processes and vary strongly between the individual E/S0s. 
Most E/S0 galaxies have negative colour gradients, meaning that the centres are reddest and  colours become bluer outwards. This could be the result of a negative gradient in age, metallicity or dust, or some combination of these.

Tamura et al. (2000) and Tamura and Ohta (2003) estimate the mean colour gradient in E/S0s as ${{\Delta(B-R)}\over{\Delta(\rm log~r)}}=-0.09$ mag $\rm dex^{-1}$ (with a scatter of 0.04), and attributed this to a radial gradient of metallicity (rather than age) 
${{d(\rm log~Z)}\over{d(\rm log~r)}}= -0.3\pm 0.1$, approximately constant to $z\simeq 1$. Wu et al. (2005) similarly estimated ${{d({\rm log}~Z)}\over{d(\rm log~r)}}= -0.25$ with a large scatter $\sigma=0.19$. Mehlert et al. (2003) found negative metallicity gradients in  Coma-cluster ellipticals, but negligible gradients in age or $\alpha$/Fe. Michard (2005) concluded colour gradients in 50 nearby ellipticals were primarily due to metallicity gradients and the effects of dust were usually small. La Barbera and Carvalho (2009) compare optical and near-IR colour gradients and again conclude they are produced by negative gradients in metallicity (they find the radial age gradients in E/S0s may be even be positive).   

Kobayashi (2004) performed detailed chemodynamical simulations of a set of over 100 model elliptical galaxies, with star-formation and merging, and predicted that spheroidals which formed monolithically, or at least with only minor mergers, would have steeper metallicity gradients (${{d({\rm log}~Z)}\over{d(\rm log~r)}}\simeq -0.3$ to -0.5) than those formed by major mergers (${{d({\rm log}~Z)}\over{d(\rm log~r)}}\simeq -0.2)$. Within the simulated galaxy set, merger history was the primary determinant (more so than mass or luminosity) of present-day metallicity/colour gradient. This might account for the observed wide distribution in colour gradients if similar numbers of spheroidals had monolithic, minor-merger or major-merger histories, and might also lead to an environmental dependence.

In this paper we focus especially on Brightest Cluster Galaxies (BCGs), of which several thousand have been identified in the SDSS. Previously, the BCGs in the SDSS have been found to follow a steeper relation of radius to luminosity (${\rm r}_{eff}\propto L$) than the other E/S0s (${\rm r}_{eff}\propto L ^{0.6}$); (Bernardi et al. 2007; Bernardi 2009). In this paper, we look for systematic differences in their CMR and radial colour gradients with respect to other E/S0 galaxies, especially those in the same luminosity range ($M_r<-22.5$ to $M_r\simeq -25$). Some models, for example,  predict a much flatter CMR for the BCGs, as a result of formation from a relatively large number of mergers of (already) old and red galaxies (de Lucia and Blaizot 2007). Ko and Im (2005) found indications that the colour gradients of ellipticals in dense cluster environments tended to be less strong than in field environments, which again could be due to (more) mergers.

In Section 2 of this paper we describe the data used and our selection of early-type galaxies and BCGs. In Sections 3 and 4 we investigate and compare their CMR and colour-$\sigma$ relation. In Section 5 we use an estimator of colour gradient based on the ratio of $g$ and $r$-band effective radii to compare the colour gradients of BCGs and other E/S0s and examine the dependences on luminosity, radius, and density. In Section 6 we look at the influence of stellar age, as estimated from the spectra. We conclude in Section 7 with summary and further discussion.

SDSS magnitudes are given in the AB system where $m_{AB}=-48.60-2.5$ log 
$F_{\nu}$ (in ergs $\rm cm^{-2}s^{-1}Hz^{-1}$); equivalently, $m_{AB}=0$ is 3631 Jy.
We assume throughout a spatially flat cosmology with $H_0=70$ km $\rm 
s^{-1}Mpc^{-1}$, $\Omega_{M}=0.27$ and $\Omega_{\Lambda}=0.73$, giving the age of the Universe as 13.88 Gyr.
\section{SDSS Data and Galaxy Sample Selection}
We first describe the selection of E/S0 and BCG galaxies used in the colour-magnitude relation analysis. For the following analysis of colour gradients, we exclude some objects to give somewhat smaller samples, as described in Section 5.1.

In Paper I we selected 70378 E and S0 galaxies out of a total of 367471 galaxies in the DR4 spectroscopic sample of the Sloan Digital Sky survey, with parameters updated to DR6. Strict criteria were applied; the E/S0s had to have de Vaucouleurs profile fractions ($\rm frac_{deV}$) from the SDSS disk-plus-bulge fits of 1.0 in both $g$ and $r$ (i.e. galaxies with any significant disk component are excluded), `eclass' spectroscopic classification parameters $<0$ (signifying absorption-line spectra), and dereddened $r$-band model magnitudes $14.5<r<17.5$. Redshift range is $0<z<0.36$. The same sample is used here, and (as in Paper I) we include the small `sky subtraction' corrections from Hyde and Bernardi (2009); applied to the SDSS photometry these have the effect of slightly increasing the effective radii and brightening the model magnitudes (in all passbands) for the galaxies of larger angular size. 

Here we make use of two samples of BCGs, both selected from the SDSS. The first are the `C4 BCGs', a subsample of the first-ranked galaxies within the catalog of clusters detected using a `C4' cluster-detection 
algorithm on the SDSS-DR2 data (Miller et al. 2005). Bernardi et al. (2007) selected a subset of 286 (of the original list of 748) BCGs as those with $M_r<-21$, velocity dispersion ($\sigma$) measurements and at least 10 other $M_r<-21$ galaxies within 1.4 Mpc. We further restrict the Bernardi et al. (2007) C4 sample to the 166 BCGs which lacked any spiral or disk component and had no neighbours close enough to confuse the radial profile. These cover the 
redshift range $z=0.03142$ to 0.16315 with a mean $z=0.0848$.

 For the C4 BCGs (only) we have improved photometry available; the galaxies were individually re-examined using software developed with 
`Matlab', which fitted new 'Galmorph' magnitudes and radii in a system equivalent to the SDSS model magnitudes (Hyde and Bernardi 2009), with more accurate sky subtraction (not requiring further corrections).

The second, larger sample of BCGs is selected from the `max-BCG' catalog of Koester et al. (2007), and previously studied by Bernardi (2009); as in Bernardi (2009) the selection criteria are to use only galaxies with SDSS spectra and which have $\rm frac_{deV}>0.8$ in both $g$ and $r$, $60<\sigma<450$ km $\rm s^{-1}$, a concentration index $>2.4$, and a dereddened model magnitude $14.5<r<17.5$. This provides a sample of 4919 BCGs  covering a deeper redshift range $z=0.04144$ to 0.31722 with a mean $z=0.1837$. For the max-BCG  galaxies we have only the standard DR6 photometry and effective radii, to which we applied the sky-subtraction corrections from Hyde and Bernardi (2009).

Fourthly, we select a comparison sample of high luminosity non-BCG E/S0s.
This is the subset of the 70378 E/S0 galaxies with absolute magnitudes $M_r<-22.5$ (there are 20110) after excluding all galaxies identified as BCGs i.e. members of either the C4 BCG or max-BCG lists, to give a non-BCG sample of 18225.
\section{Colour-Magnitude Relation}
\subsection{Background}
Bernardi et al. (2003 and 2005) investigated the CMR of early-type galaxies in the SDSS, deriving  the rest-frame $g-r$ colours from the observer-frame photometry using k-corrections calculated from models and template spectra (e.g. Hogg 2002). In Roche, Bernardi and Hyde (2009, hereafter Paper I), we investigated the CMR of a larger sample of 70378 E/S0 galaxies from later SDSS releases (Adelman and McCarthy et al. 2006, 2008), again in terms of the mean rest-frame $g-r$ colour as a function of red absolute magnitude ($M_r$). This time, rather than using modelled k-corrections, we obtained rest-frame colours using the observed SDSS spectra of the individual galaxies (independently of any models), using two methods described briefly below (see Paper I for more details).

 Firstly, we extracted rest-frame $g$ and $r$ magnitudes directly from the spectra by integrating directly over the relevant wavelength ranges. We term these spectra-derived colours. This method does not use the imaging data at all or require any k-correction. However, the spectra-derived colour only measures $g-r$ in the central 3 arcsec diameter aperture covered by the spectrograph fibre. As the angular size of an SDSS galaxy may be much larger than this, and will be correlated with luminosity, this can introduce a bias into the CMR (e.g. changing its slope) if the galaxies have radial colour gradients.

 The second method was to  integrate the  spectra over the observed and rest-frame passbands; the difference of these two magnitudes provided a spectra-derived k-correction for each individual galaxy. The SDSS imaging photometry provides magnitudes based on profile fits to the galaxies (in the case of these galaxies, de Vaucouleurs profiles), with the effective radius fixed across all passbands to that $r_{eff}$ measured by fitting in the $r$-band. These are termed `model' magnitudes and colours. 
Subtracting the spectra-derived k-corrections from the model magnitudes will then give rest-frame colours for the 
integrated emission from the whole of each galaxy. However, the model magnitude colours can also be subject to biases - in the opposite direction - because a k-correction based on an fixed-aperture spectrum is being applied to a colour derived from a model fit, which may be from a much larger aperture.

In agreement with previous studies, we found approximately linear CMRs, redder $g-r$ colours for the more luminous galaxies, and a mild blueward shift of the CMR with increasing redshift. However, our CMR based on spectra-derived colours was steeper than the CMR from model-magnitude colours (${{d (g-r)}\over{d M_r}}\simeq -0.026$ compared to -0.018), and showed more evolution with redshift. Similar variation in the CMR slopes between aperture and `total' photometric systems had been reported on by Scordeggio (2001) and Bernardi et al. (2003). The colour differences $(g-r)_{spec}-(g-r)_{model}$ and $(g-r)_{fib}-(g-r)_{model}$ will be sensitive to radial colour gradients.

In Paper I we also examined the relation of colour to internal stellar velocity dispersion $\sigma$, the $C\sigma R$. As expected, galaxies with greater $\sigma$ were redder, but we did not see a difference in slope between the spectra-derived and model-magnitude based $C\sigma R$; both gave ${{d(g-r)}\over {d({\rm log}~\sigma)}}\simeq 0.19$. 
 This suggested that colour gradients might depend differently on luminosity and on $\sigma$, and might have a positive correlation only with luminosity. We found that the difference between the 3-arcsec aperture and model magnitude  colours was correlated with the 
residual of $\sigma$ relative to the mean $\sigma$--$M_r$ relation, and similarly with the $M_r$ residual to the $M_r$--$\sigma$ relation. These two correlations indicated a stronger (negative) colour gradient was associated with a higher luminosity and/or a lower $\sigma$ relative to the mean $\langle\sigma|M_r\rangle $. We investigate this further in Section 5.2 below, after again looking at CMRs.

In this paper we again make use of rest-frame $g-r$ colours and absolute magnitudes $M_r$ obtained from the SDSS model magnitudes (or the Galmorph magnitudes in the case of the C4 BCGs), corrected to rest-frame using the spectra-derived k-corrections calculated as described in Paper I (including the `signal/noise' correction, which had the effect of reducing the $g$-band k-correction by a few hundredths of a magnitude for the fainter objects, on the basis of an apparent discrepancy between the spectral calibration and the imaging photometry). We also make use of the rest-frame colours derived directly from the spectra. 
\subsection{The CMRs of BCGs compared with other early-type galaxies} 
Figure 1 shows the CMR, in model-magnitude $g-r$ colours k-corrected to rest-frame and averaged in $\Delta(M_r)=0.5$ intervals, for the full E/S0 sample of Paper I (divided into 7 redshift intervals), the C4 BCGs and the max-BCGs.  As in Paper I the absolute magnitudes are `de-evolved' by adding $+0.86z$ magnitudes to separate the colour evolution from the effect of luminosity evolution.
\begin{figure} 
\includegraphics[width=0.7\hsize,angle=-90]{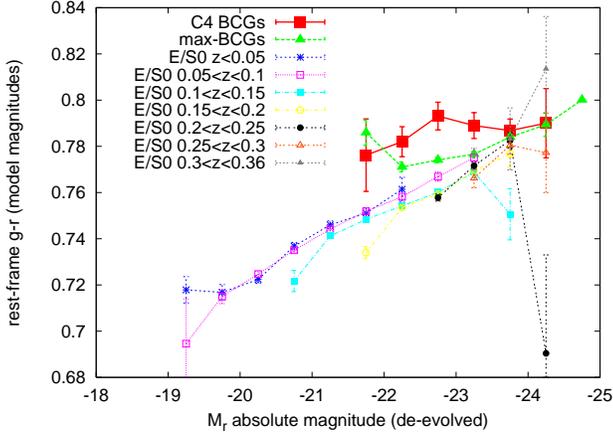} 
\caption{Colour-magnitude relation, with {\tt model}-magnitude colours averaged in $\Delta(M_r)=0.5$ intervals, of Brightest Cluster Galaxies (BCGs), separately for the  lower-redshift C4 and the higher-redshift max-BCG samples, with the full sample of E/S0s divided by redshift.}
 \end{figure}

The C4 and max-BCG galaxies have similar CMRs, with the lower redshift C4s being slightly ($\sim 0.01$ mag) redder. Both sets of BCGs are on average slightly redder than the E/S0s at a similar luminosity and redshift and their CMRs are also flatter in slope. 

As in Paper, I we fit the galaxy CMRs with the form
\begin{equation}
 \langle (g-r)_{rf}|M_r,z\rangle = a_0 + a_1\,M_r +a_2\,z
 \label{cmrfit}
\end{equation}
with separate linear dependencies on $M_r$ ($a_1$, the slope) and $z$ ($a_2$, the evolution) and no cross-term.
 The first line of Table 1 gives the fitted parameters for the (model-magnitudes) CMR of our full E/S0 sample; these are the same as given in Paper I. 

Figure 2 compares the CMRs of the max-BCG and the $M_r<-22.5$ non-BCG galaxies, divided into two intervals of redshift, and the C4 BCGs (in a single redshift interval -- all are at $z<0.18$); this is shown for both model-magnitude and spectra-derived colours.
\begin{figure} 
\includegraphics[width=0.7\hsize,angle=-90]{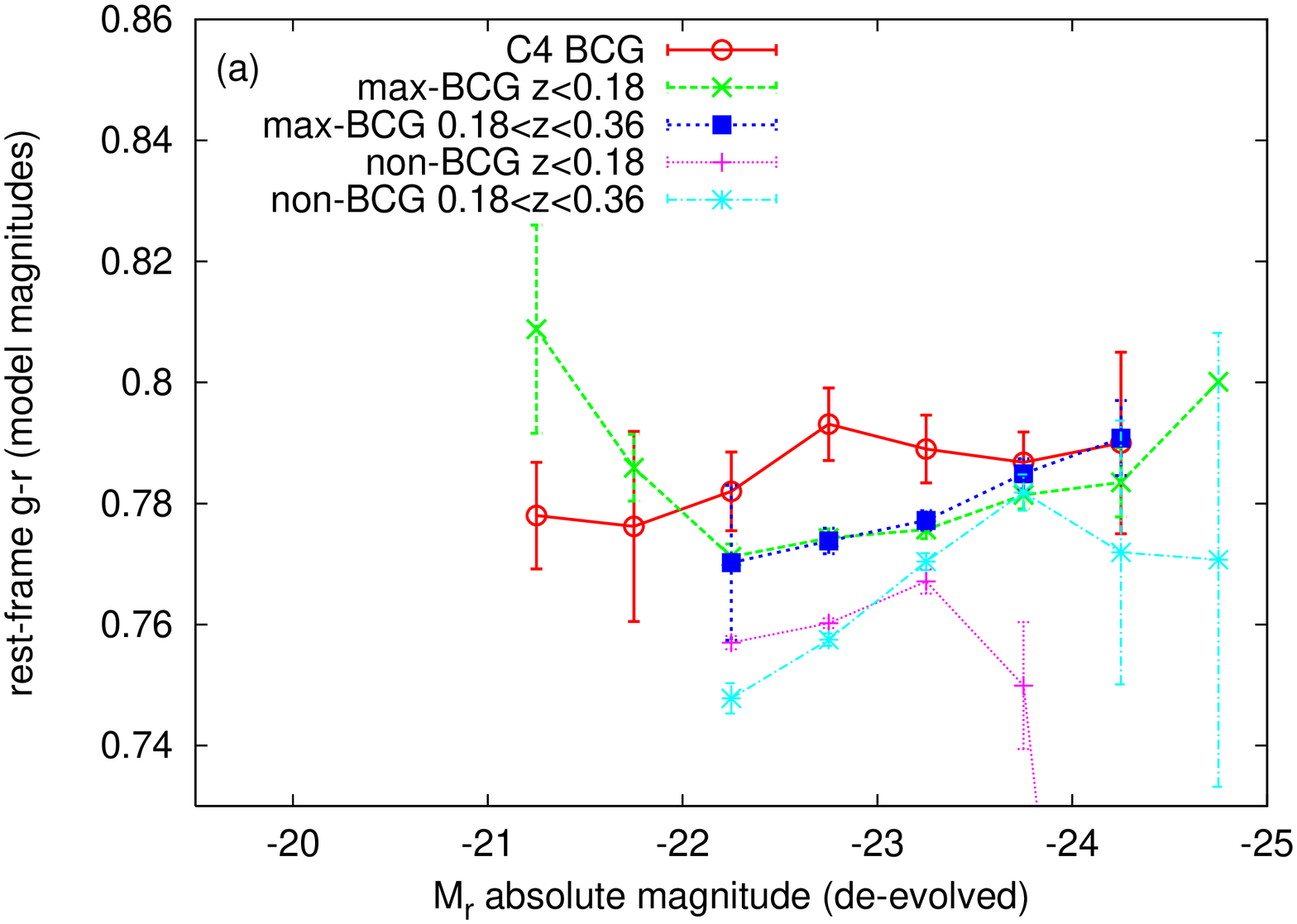} 
\includegraphics[width=0.7\hsize,angle=-90]{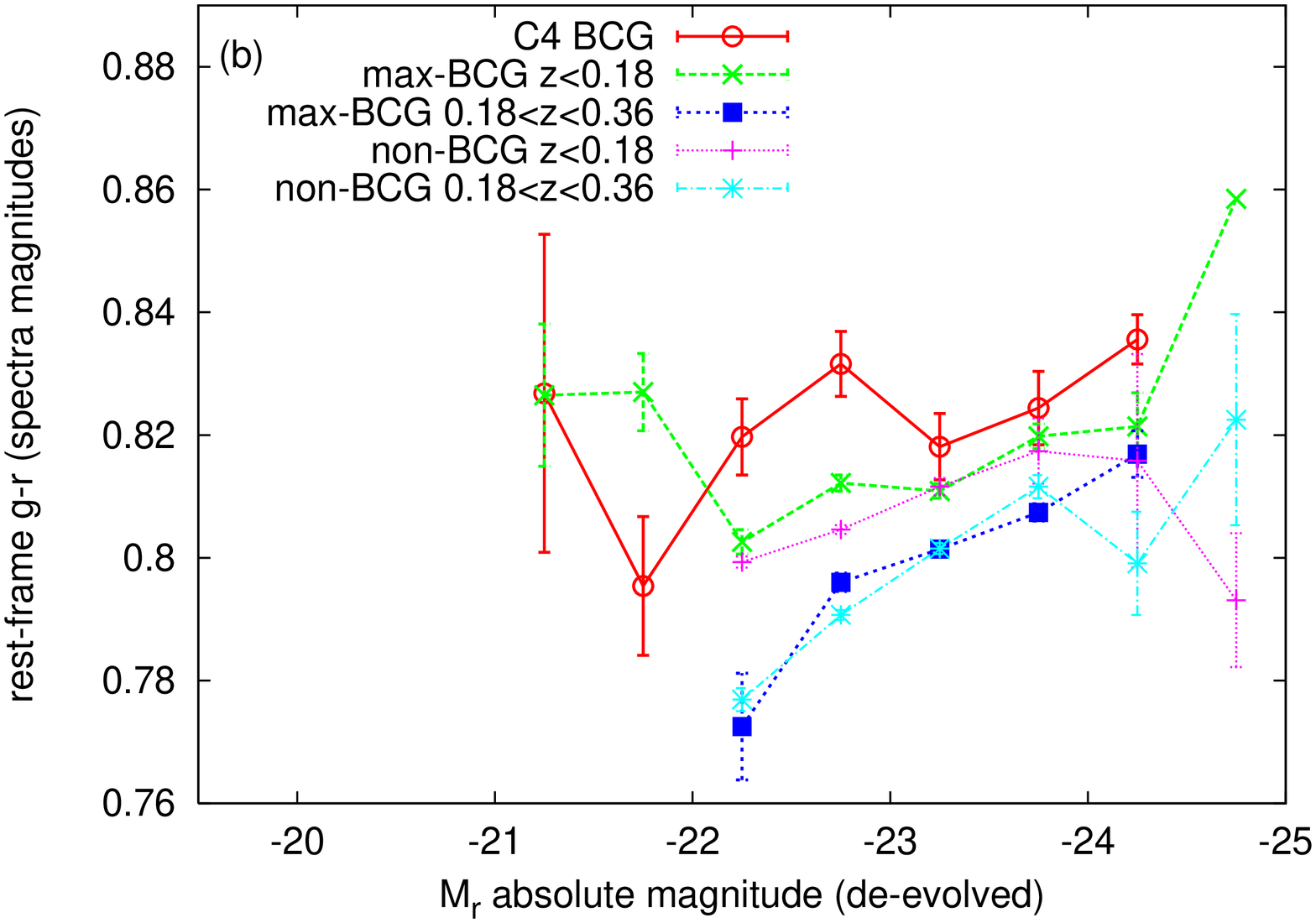}
\caption{Colour-magnitude relation, in (a) model magnitudes, (b) spectra-derived magnitudes, for the max-BCGs, the sample of 18225 $M_r<-22.5$ non-BCG E/S0s (divided into two intervals of redshift) and the C4 BCGs.} 
 \end{figure}
The two BCG samples have quite similar CMRs. In model magnitudes, the non-BCG CMR is significantly bluer than that of either BCG sample (in each redshift interval), except perhaps at the highest luminosities, and appears steeper in slope. In spectra-derived magnitudes, all the CMRs are moved slightly redwards  (this could be the effect of negative colour gradients), but there is much less  colour difference in between the BCGs and non-BCGs

Table 1 gives the parameters fitted to these CMRs (an evolution term is only fitted for the max-BCG sample of BCGs as the C4 sample is too small and covers a narrow redshift range). These indicate that the CMR of (both sets of) BCGs, although not entirely flat, has a slope ($a_1$) less than half that of the non-BCG E/S0s of similar luminosity. This difference in $a_1$ is seen for both model and spectra-derived colours.
 In model magnitudes we see little or no colour evolution ($a_2$), but in spectra-derived magnitudes there is
(as noted in Paper I) some blueward evolution with redshift. This colour evolution is seen here for both BCGs and non-BCGs, with marginal evidence of a slightly lower rate for the BCGs.
 
The mean residual of the max-BCG $g-r$, relative to to the $\langle g-r | M_r,z\rangle$ fitted to the non-BCG $M_r<-22.5$ E/S0s, is $0.0090\pm0.0007$ in model magnitudes and $0.0016\pm 0.0005$ in spectra-derived magnitude. For the smaller C4 sample these $\Delta(g-r)$ are $0.0208\pm 0.0030$ and $0.0023\pm 0.0029$ magnitudes. Thus the BCGs are significantly redder than other E/S0 (matched in luminosity) in their mean model-fit colours but with much less or no difference in their spectra-derived (i.e. central 3 arcsec aperture) colours.  

\onecolumn
\begin{table}
\begin{tabular}{lccc}
\hline
Sample & $a_0$ & $a_1$ (slope) & $a_2$ (evolution) \\
\hline
\smallskip
All E/S0 (model mag) & $0.3794\pm 0.0093$ & $-(0.0174\pm 0.0005)$ &  $-(0.0743\pm 0.0075)$ \\
C4 BCGs (model mag) & $0.6604\pm 0.1084$ & $-(0.0056\pm 0.0047)$ & - \\
max-BCGs (model mag) & $0.5877\pm 0.0403$ & $-(0.0082\pm 0.0018)$ & $-(0.0040\pm 0.0172)$ \\
non-BCG (model mag) & $0.3132\pm 0.0374$ & $ -(0.0200\pm 0.0017)$ & $-(0.0402\pm 0.0133)$ \\
C4 BCG (spectra mag) & $0.7121\pm 0.1028$ &  $-(0.0049\pm 0.0045)$ & - \\
max-BCGs (spectra mag) & $0.5118\pm 0.0282$ & $-(0.0142\pm 0.0013)$ & $-(0.1825\pm 0.0121)$ \\
non-BCG (spectra mag) & $0.2234\pm 0.0286$ & $-(0.0270\pm 0.0013)$ & $-(0.2326\pm 0.0104)$ \\
\hline
\end{tabular}
\caption{Fits to colour-magnitude relations}
\end{table}

\begin{table}
\begin{tabular}{lccc}
\hline
Sample & $a_0$ & $a_1$ (slope) & $a_2$ (evolution) \\
\hline
\smallskip
All E/S0 (model mag) & $0.3285\pm 0.0044$ & $0.1914\pm 0.0021$ & $-(0.1289\pm 0.0053)$ \\
C4 BCGs (model mag) & $0.5610\pm 0.0859$ & $0.0946\pm 0.0359$ & - \\
max-BCGs (model mag) & $0.6584\pm 0.0260$ & $0.0474\pm 0.0109$ & $0.02049\pm 0.01528$ \\
non-BCG (model mag) & $0.2270\pm 0.0137$ & $0.2288\pm 0.0059$ & $-(0.0664\pm 0.0114)$\\
C4 BCG (spectra mag) & $0.6658\pm 0.082$ & $0.0656\pm 0.0343$ & - \\
max-BCGs (spectra mag) & $0.6538\pm 0.0182$ & $0.0733\pm 0.0077$ & $-(0.1375\pm 0.0100)$ \\
non-BCG (spectra mag) & $0.4381\pm 0.0108$ & $0.1661\pm 0.0046$ & $-(0.1985\pm 0.0090)$ \\
\hline
\end{tabular}
\caption{Fits to colour-$\sigma$ relations}
\end{table}
\twocolumn
\section{Colour-Velocity Dispersion Relation}
We can make the same comparison for the relation of colour to internal stellar velocity dispersion ($\sigma$), hereafter the C$\sigma$R. As described in Paper I, we apply small corrections to the SDSS $\sigma$ measurements to convert from the spectrograph aperture to a physical scale of $\rm r_{eff}/8$, assuming the $\sigma\propto\rm  r^{-0.04}$  of J{\"o}rgensen et al. (1995).

 Figure 3 shows the colour-$\sigma$ relation (C$\sigma$R), defined in model magnitude rest-frame $g-r$, for the full E/S0 sample and the two BCG samples. As in the CMR, we see the BCGs tend to be slightly redder (at a given $\sigma$) and have a flatter C$\sigma$R.
\begin{figure} 
\includegraphics[width=0.7\hsize,angle=-90]{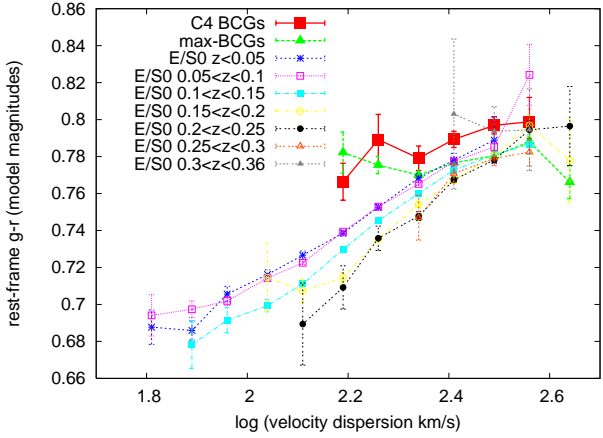} 
\caption{Colour-$\sigma$ relation with model-magnitude colours, of Brightest Cluster Galaxies (BCGs), separately for the  lower-redshift C4 and the higher-redshift max-BCG samples, with the full sample of E/S0s divided by redshift.} 
 \end{figure}
We fit with 
\begin{equation}
 \langle (g-r)_{rf}|\sigma,z\rangle = a_0 + a_1\,{\rm log}~\sigma +a_2\,z
 \label{csrfit}
\end{equation}
The first line of Table 2 gives the fitted parameters for the whole E/S0 sample (as given in Paper I).
Figure 4 compares the C$\sigma$R of the two BCG samples with the non-BCG $M_r<-22.5$ galaxies, computed for both model and spectra-derived magnitudes. Table 2 gives the fitted parameters (again, evolution is not fitted for the small C4 sample).
\begin{figure}
\includegraphics[width=0.7\hsize,angle=-90]{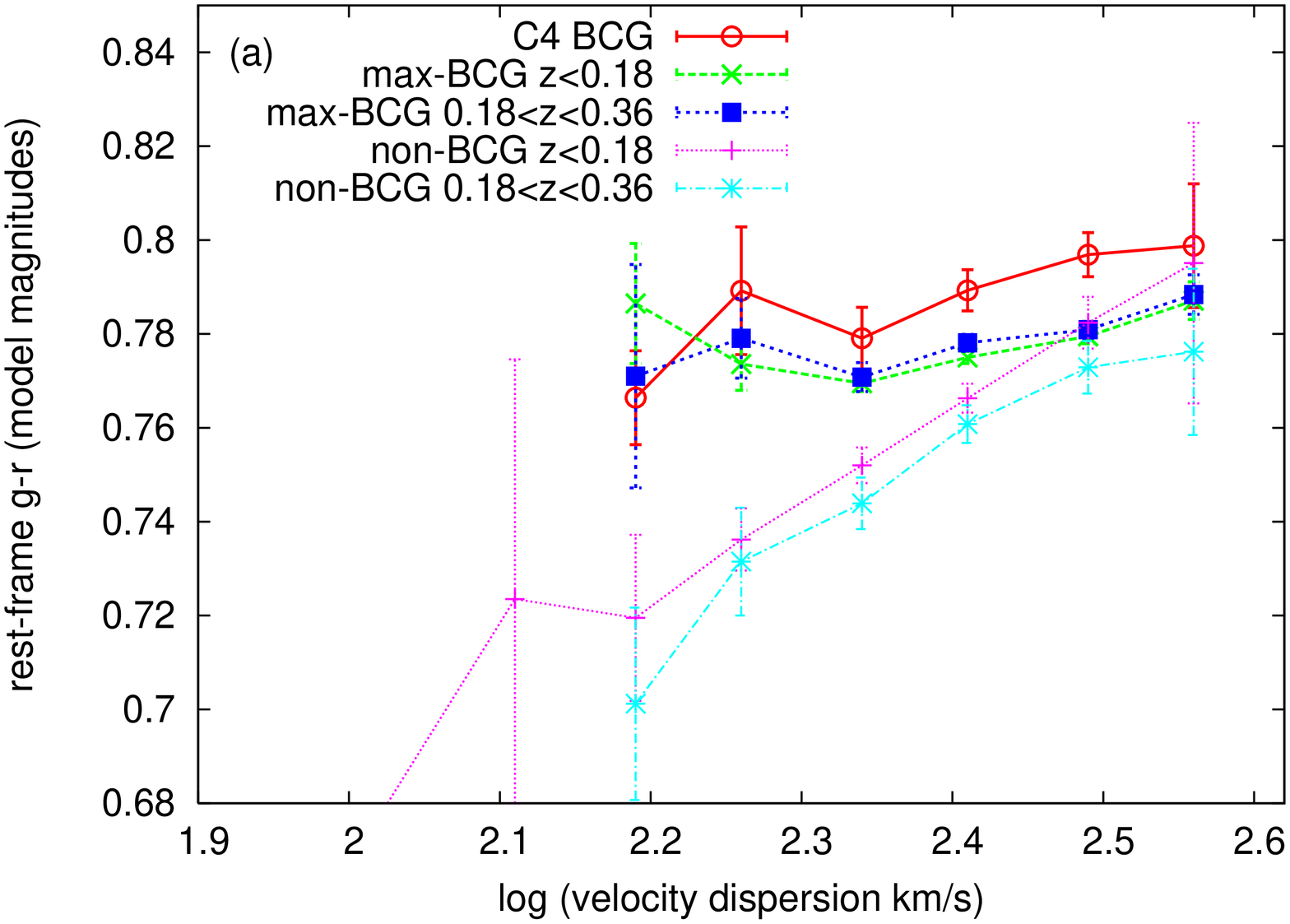}
\includegraphics[width=0.7\hsize,angle=-90]{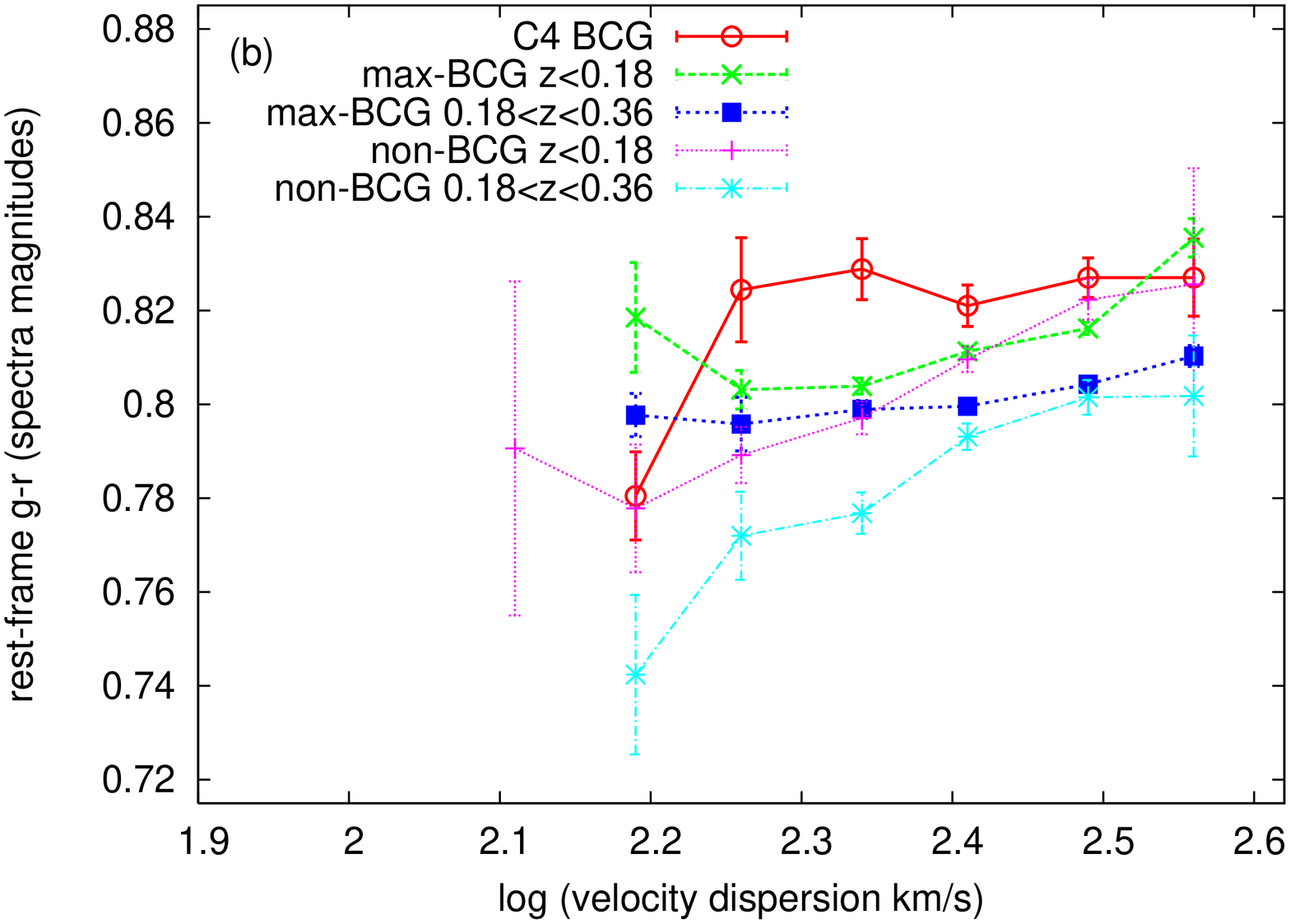} 
\caption{Colour-$\sigma$ relation of the C4 BCGs and max-BCGs, compared with the 18225 $M_r<-22.5$ non-BCG E/S0s. Shown for (a) model and (b)spectra-derived magnitudes.} 
 \end{figure}
Again the BCG ${{d (g-r)}\over{d{\rm log}~\sigma}}$ slope is less than half that of the non-BCG E/S0s. As we saw for the CMR, the spectra-derived colours show more blueward evolution than the model-magnitude colours, with the BCGs having a slightly lower evolution rate.

 In the C$\sigma$R the mean colour residual of the max-BCGs relative to the non-BCG $\langle g-r|\sigma,z\rangle$ is 
$0.0070\pm 0.0007$ mag in model-magnitude $g-r$ and $0.0016\pm 0.0005$ mag in spectra-derived colour, and for the C4 sample these residuals are $0.0187\pm 0.0029$ and $0.0042\pm 0.0028$ mag; very similar to the colour offsets in the CMR. 

\section{Colour Gradients in E/S0s and BCGs}
\subsection{Method of Colour Gradient Analysis}
As a simple and rapidly computable quantifier of radial colour gradient, we make use of the ratio of the de Vaucouleurs effective radii fitted in the $g$ band and the $r$ band. This ratio will be unity for a galaxy with no colour gradient and $>1$ for a negative colour gradient, i.e. with the centre redder than the outer annuli. A ratio of de Vaucouleurs $r_{eff}$ (in $V$ and $K$)  was previously used as a colour gradient measure by Ko and Im (2005).

To be redshift-independent, we need to use ${\rm r}_{eff}(g)$ and ${\rm r}_{eff}(r)$ as they would be measured in the the rest-frame. We estimate these by linear interpolation between the ${\rm r}_{eff}$ fitted in the four passbands $griz$, using in the interpolation  the passband mean wavelengths 4680, 6180, 7500, 8870$\rm \AA$.

We adopt a colour gradient measure as the restframe-corrected ${{{\rm r}_{eff}(g)}\over{{\rm r}_{eff}(r)}}-1$. This will be related to but not directly equivalent to ${{d(g-r)}\over{d(\rm log~r)}}$, a measure of colour gradient obtained by directly fitting to the colour profiles of each galaxy. The relation between the two measures will depend on the exact form of the intensity profile and its variation with wavelength, and this may vary between individual ellipticals. To estimate this relation for the SDSS sample and to verify whether  ${{{\rm r}_{eff}(g)}\over{{\rm r}_{eff}(r)}}-1$ is a valid measure of colour gradient across this range of galaxies, we obtain from the SDSS Data Release 7 the tabulated $g$ and $r$ profiles for a small sample (90 galaxies).

For the 90 galaxies, which include 21 BCGs, we make a linear error-weighted fit of the form
\begin{equation}
 (g-r)=a_0+a_1~(\rm log~radius)
\end{equation}
 to the $g-r$ colour profile over all annular bins from an inner radius 0.67 arcsec to the outer limit of the tabulated profile, usually 28 arcsec, examining each fit by eye.

 Figure 5 shows the gradient  ${{d(g-r)}\over{d(\rm log~r)}}$ plotted against our colour gradient  ${{{\rm r}_{eff}(g)}\over{{\rm r}_{eff}(r)}}-1$. It is clearly correlated with a linear correlation coefficient 0.768 and best fit relation $-0.0119(\pm 0.0038)-0.9398(\pm 0.0336)x$, passing very close to the origin as expected, but the scatter of 0.122 is rather large. This is in part because a significant minority of the ellipticals have colour profiles not linear in log radius; we see non-monotonic gradients for about $10\%$ and in these cases, the $r_{eff}$ ratio tends to be sensitive to the colour profile at a larger radius than the direct fit (which is weighted towards smaller radii where the error bars are smaller).
\begin{figure}
\includegraphics[width=0.7\hsize,angle=-90]{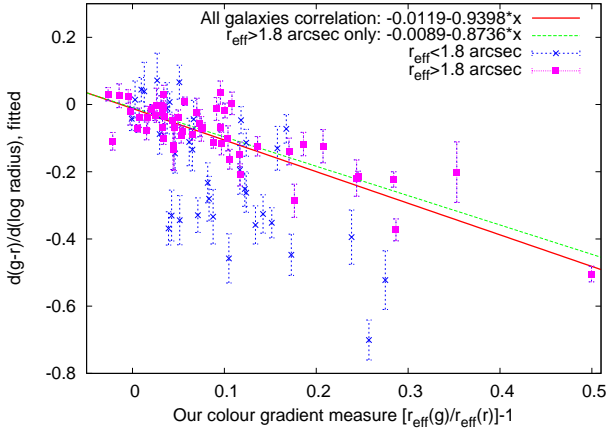} 
\caption{The relation between two measures of colour gradient for an early-type  galaxy,  the ratio of de Vaucouleurs radii ${{{\rm r}_{eff}(g)}\over {{\rm r}_{eff}(r)}}-1$, used in this paper, and  ${{d(g-r)}\over{d(\rm log~r)}}$ from a fit to the colour profile, for 90 galaxies from our sample with symbol types indicating apparent size (in the $r$ band). Best-fit linear correlations are plotted for all 90 and for the 51 galaxies with
 $r_{eff}>1.8$ arcsec.} 
\end{figure}
\begin{figure}
\includegraphics[width=0.7\hsize,angle=-90]{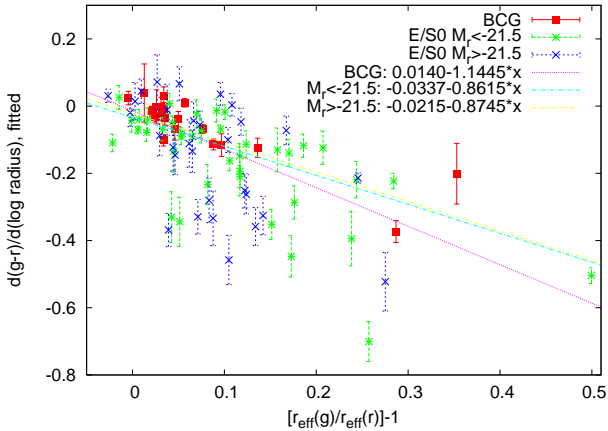}
\caption{As Figure 5 but with symbols indicating that galaxies are 
 BCGs, higher luminosity ($M_r<-21.5$) but non-BCG E/S0s, or lower lminosity. Best-fit linear correlations are shown for each of these three classes. These correlations are quite similar but the scatter is less for the BCGs.}
\end{figure}

 Despite this, BCGs and high and low luminosity E/S0s, considered separately (Figure 6), show similar relations between the two colour gradient measures. However, the relation is much tighter for the BCGs with a scatter of only 0.053, compared to 0.123 (high L E/S0) and 0.150 (low L E/S0). As a further test, we examine 
 (Figure 7) the residuals of the 
${{d(g-r)}\over{d(\rm log~r)}}$ to the best-fit linear relation, as a function of $M_r$ and apparent $r_{eff}$ (in the $r$-band); a significant trend could introduce a bias into the investigations of this paper. 

We find the scatter tends to increase at lower luminosity but there is not an obvious systematic trend. Neither do we see a trend with apparent radius, down to about 1.8 arcsec. However, for galaxies smaller than this radius the scatter increases greatly and asymmetrically, which is probably unacceptable for our purposes. For galaxies with $r_{eff}<1.8$ arcsec it may not be possible to meaningfully characterize the colour gradient at all using SDSS data. For early-type galaxies of larger apparent size, including the BCGs, the ratio of de Vaucoleurs radii does appear to be a valid, consistent and approximately linear colour gradient measure.

Returning to Figure 5, if all $r_{eff}<1.8$ arcsec galaxies are excluded, leaving 51, the best-fit linear relation changes only slightly to $-0.0089(\pm 0.0039) -0.8736(\pm 0.0346)x$, but the scatter is decreased by more than a factor of two to a much more acceptable 0.056 (similar to the scatter for BCGs only), and the correlation coefficient is increased to 0.862.
\begin{figure}
\includegraphics[width=0.7\hsize,angle=-90]{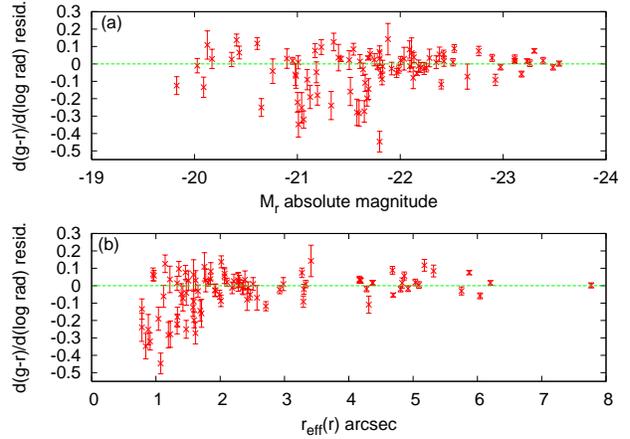}
\caption{Residuals of fitted gradient  ${{d(g-r)}\over{d(\rm log~r)}}$ to the best-fit relation $-0.0119-0.9398[{{{\rm r}_{eff}(g)}\over {{\rm r}_{eff}(r)}}-1]$, plotted against (a) absolute magnitude $M_r$ and (b) apparent size $r_{eff}$.}
\end{figure}

Some galaxies in our sample have very strong ($>0.5$) or inverted ($<0$) colour gradients (i.e. ${\rm r}_{eff}(g)<{\rm r}_{eff}(r)$). Examining some of these, we find that some are  galaxies confused with or very close to (red or blue) stars, and that there are also  some misclassified spirals present, despite our strict selection criterion of $frac_{Dev}=1$ in both $g$ and $r$. To reduce any biases that these types of objects may introduce, we exclude some galaxies from the colour gradient analysis. 

 Firstly, we excluded the objects with gradients (i.e. ${{{\rm r}_{eff}(g)}\over {{\rm r}_{eff}(r)}}-1$) calculated as $<-0.5$ or $>1$, far out of the normal range (there were only 130/70378).
Secondly, we examined by eye all galaxies in the E/S0 sample with gradients 
0.4--1.0; these numbered 1869 out of 70378. It was estimated that 108 of these were elliptical-elliptical mergers, 26 were mergers involving spirals, 98 were barred spirals, 142 were edge-on disks, 120 were other galaxies with visible spiral arms and 88 were confused with stars. All of these except  the elliptical mergers, 474 galaxies, were marked for exclusion. It should also be noted that many other high colour gradient galaxies were asymmetric or disturbed.

 At lower colour gradients the proportion of face-on spirals decreases steeply ($13\%$ at gradient $>0.4$, $5\%$ at 0.3, $2\%$ at 0.2) but 
less steeply for edge-on disks ($8\%$ at gradient $>0.4$, $8\%$ at 0.3, $3\%$ at 0.2). However, the latter could easily be removed as they had axis ratios (from the $r$-band de Vaucouleurs fits) of typically $b/a\simeq 0.3$ and always $b/a<0.4$, whereas virtually all ellipticals had $b/a>0.4$. We therefore exclude all galaxies with $b/a<0.4$, whatever their colour gradient.
Lastly, because of their very large scatter on Figure 7, we exclude all objects with $r$-band $r_{eff}$ smaller than 1.8 arcsec.

 These exclusion criteria, for the purposes of colour gradient analysis,  reduce the all E/S0 sample to 37001, the $M-r<-22.5$ non-BCG sample to 13652, and the max-BCG sample to 4485 (we note that these exclusions make very little difference to the colour-magnitude relations of the respective samples, the main effect being increased error bars at the lowest luminosities).
\subsection{Results}

After this filtering, for the whole E/S0 sample the mean ${{{\rm r}_{eff}(g)}\over{{\rm r}_{eff}(r)}}-1$ is $0.11115\pm 0.00061$ with a scatter 0.11724. From the relation on Figure 6 this corresponds to  ${{d(g-r)}\over{d({\rm log~r)}}}\simeq -0.10$ and so lies half-way between the mean gradient ${{d(g-r)}\over{d({\rm log~r)}}}=-0.071$ measured for similarly selected E/S0s by La Barbera and Carvalho (2009) and the larger ${{d(g-r)}\over{d({\rm log~r)}}}=-0.14$ found by Suh et al. (2010) for `red-cored' morphological early-types after the exclusion of zero and inverted gradients. 

 For the $M_r<-22.5$ non-BCGs sample the mean gradient is similar, $0.10550\pm 0.00098$ with a scatter 0.11414. But for the max-BCGs the mean gradient is significantly lower at $0.08142\pm 0.00114$ with scatter 0.07645. Hence, we find evidence that BCGs have flatter colour gradients than other E/S0s; the difference is modest but with this large sample highly significant - the BCGs' mean gradient is $22.8\pm 1.8\%$ less than for the $M_r<-22.5$ non-BCGs and  $26.7\pm 1.8\%$ less than for all E/S0.

Figure 8 shows the mean ${{{\rm r}_{eff}(g)}\over{{\rm r}_{eff}(r)}}-1$ as a function of absolute magnitude $M_r$. The luminosity dependence is mild. There is a broad maximum at $M_r\simeq -22$, with a decrease of up to 
$20\%$ to higher luminosity (BCGs excluded) and $24\%$ to lower luminosities. BCGs have a significantly lower mean colour gradient than other E/S0s across their range of luminosities. 
\begin{figure} 
\includegraphics[width=0.7\hsize,angle=-90]{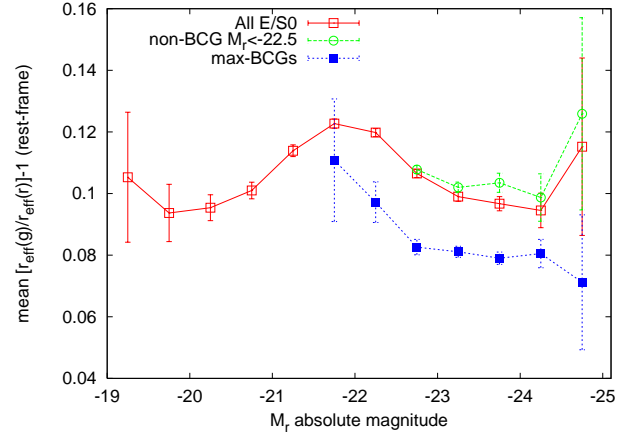} 
\caption{The mean ${{{\rm r}_{eff}(g)}\over{{\rm r}_{eff}(r)}}-1$ (corrected to rest-frame), a measure of colour gradient, for the full E/S0 sample, the $M_r<-22.5$ non-BCGs and the max-BCGs, as a function of absolute magnitude $M_r$.} 
 \end{figure}
Figure 9 shows the mean ${{{\rm r}_{eff}(g)}\over{{\rm r}_{eff}(r)}}-1$ with the galaxies divided by velocity dispersion $\sigma$. Colour gradient is maximum at low velocity dispersions of 100--150 km $\rm s^{-1}$ and decreases by almost 1/2 to $\sigma\simeq 300$ km $\rm s^{-1}$. The BCGs have a lower colour gradient that non-BCGs in each $\sigma$ interval, although the gap between the two is narrower ($\sim 0.01$) than with the galaxies in $M_r$ intervals (0.02--0.03). In the non-BCG sample, which has a sharp lower cutoff in luminosity, there is a steep increase in colour gradient at $\sigma<180$ km $\rm s^{-1}$. This reflects the finding in Paper I that a low $\sigma$ relative to luminosity is associated with a strong colour gradient. 

This is again seen on Figure 10 where the E/S0 sample is divided by absolute magnitude. Strong ($>0.15$) colour gradients are a feature of galaxies with moderate or high luminosity of $M_r<-21$ combined with log $\sigma<2.25$ ($\sigma<180$ km $\rm s^{-1}$).
\begin{figure} 
\includegraphics[width=0.7\hsize,angle=-90]{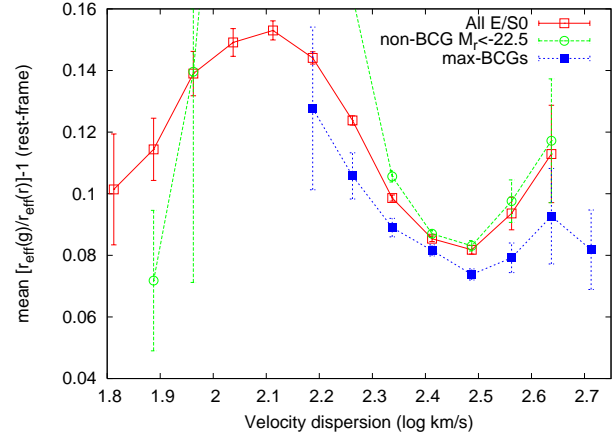} 
\caption{The mean ${{{\rm r}_{eff}(g)}\over{{\rm r}_{eff}(r)}}-1$ ratio (corrected to rest-frame), a measure of colour gradient, for the full E/S0 sample, the $M_r<-22.5$ non-BCGs and the max-BCGs, as a function of velocity dispersion $\sigma$.} 
 \end{figure}
\begin{figure} 
\includegraphics[width=0.7\hsize,angle=-90]{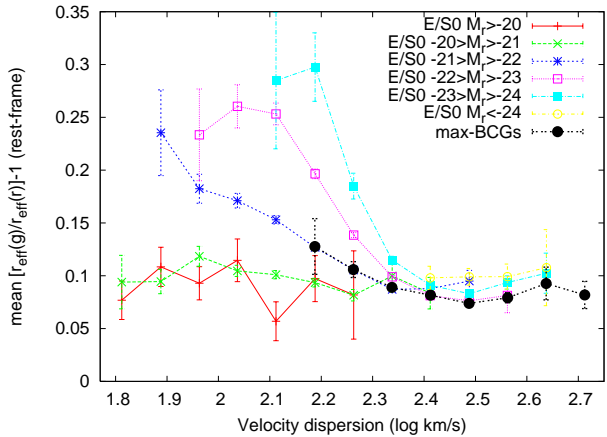} 
\caption{The mean ${{{\rm r}_{eff}(g)}\over{{\rm r}_{eff}(r)}}-1$ ratio (corrected to rest-frame) for the full E/S0 sample divided by absolute magnitude $M_r$, as a function of velocity dispersion $\sigma$.} 
 \end{figure}

Figure 11 shows colour gradient against effective radius. For the E/S0s the mean $({{{\rm r}_{eff}(g)}\over{{\rm r}_{eff}(r)}}-1$ almost doubles from the smallest radii up to $\rm r_{eff}=8$--10 kpc, and similarly for the $M_r<-22.5$ non-BCGs, for which the gradient peaks at larger radii of 11--14 kpc (presumably because of the greater mean luminosity of this sample). In both samples, mean gradient decreases at even larger radii. In BCGs the colour gradient shows a much weaker variation with radius and remains lower than in non-BCGs, especially at $5<\rm r_{eff}<20$ kpc.
\begin{figure} 
\includegraphics[width=0.7\hsize,angle=-90]{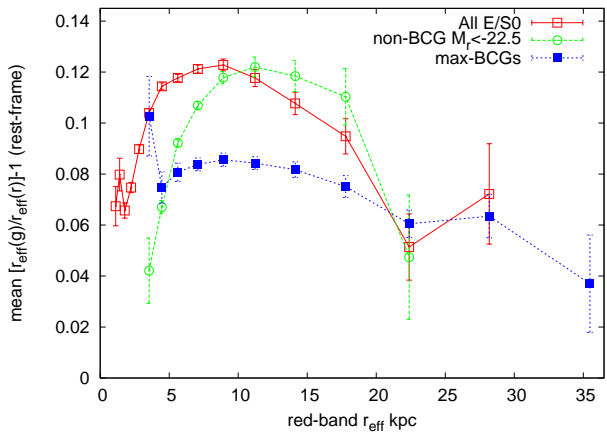} 
\caption{The mean ${{{\rm r}_{eff}(g)}\over{{\rm r}_{eff}(r)}}-1$ (corrected to rest-frame), a measure of colour gradient, as a function of red-band effective radius $r_{eff}$ for the full E/S0 sample, the $M_r<-22.5$ non-BCGs and the max-BCGs.} 
 \end{figure}

To better understand the non-monotonic variation of ${{{\rm r}_{eff}(g)}\over{{\rm r}_{eff}(r)}}-1$ with $\rm r_{eff}(r)$, and remove possible selection effects (from the flux limit of the sample) we re-examine this relation with the E/S0s divided by $M_r$ (Figure 12). For moderate luminosities there is simply a steep increase in colour gradient  with radius. The turnover at large radii is produced by $M_r<-23$ galaxies, in which colour gradient is `suppressed' and the correlation with radius is flattened. This effect is also seen, even more strongly, in the BCGs.
\begin{figure} 
\includegraphics[width=0.7\hsize,angle=-90]{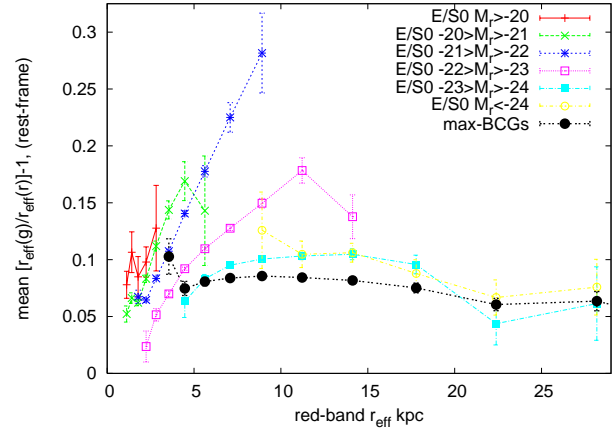} 
\caption{The mean ${{{\rm r}_{eff}(g)}\over{{\rm r}_{eff}(r)}}-1$ (corrected to rest-frame) for the full E/S0 sample divided by $M_r$, and for the max-BCGs, as a function of red-band effective radius $\rm r_{eff}$.} 
 \end{figure}

In Paper I, by comparing model and aperture colours, we found evidence that stronger colour gradients were associated with E/S0 galaxies  with high luminosity relative to $\sigma$. Figure 13  plots ${{{\rm r}_{eff}(g)}\over{{\rm r}_{eff}(r)}}-1$ against $10~{\rm log}~\sigma+M_r$ (approximately the $\sigma$ residual relative to the $\sigma$--$M_r$ relation; Forbes and Ponman 1999). In the full E/S0 sample the colour gradient increases by a factor of two from the  highest to the lowest ratios of $\sigma$ to luminosity. In the $M_r<-22.5$ non-BCG sample this increase is an even larger factor of three.  In marked contrast, the colour gradient in BCGs shows little or no dependence on the $\sigma$ residual and  remains at $\simeq 0.08$.
\begin{figure} 
\includegraphics[width=0.7\hsize,angle=-90]{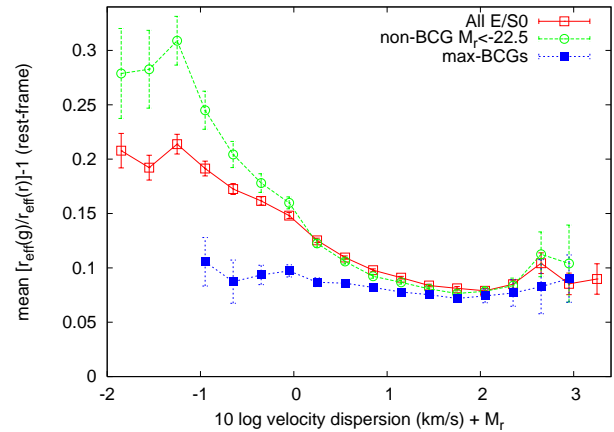} 
\caption{The mean ${{{\rm r}_{eff}(g)}\over{{\rm r}_{eff}(r)}}-1$ (corrected to rest-frame), for the full E/S0 sample, the $M_r<-22.5$ non-BCGs, and the max-BCGs, as a function of  $10~{\rm log}~\sigma+M_r$.} 
 \end{figure}

Figure 14 shows color gradient against $\sigma^2/\rm r_{eff}^2$, a more physical quantity as it is a direct measure of mass density (subject to a mild dependence on long-axis orientation). We express this in the form log ($\sigma^2/\rm r_{eff}^2$) for $\sigma$ in km $\rm s^{-1}$ and $\rm r_{eff}$ in kpc. In these units the mean density for the E/S0s is 3.38 with a scatter 0.42. This can also be converted to units of solar masses $M_{\odot}$.  From Boylan-Kolchin, Ma and Quataert (2006) the mass within $\rm r<r_{eff}$ is approximately $2.90 {{\sigma^2 \rm r_{eff}}\over{G}}$, where the gravitational constant $G=4.30\times10^{-3}$ pc $\rm M_{\odot}^{-1}$ km $\rm s^{-1})^2$. Within $r_{eff}$ the volume is ${4\over 3}\pi \rm r_{eff}^3$, and the mean mass density,
${\rm log}~(\rm M_{\odot}/kpc^3)={\rm log}~(\sigma^2/\rm r_{eff}^2)+5.21$. 
\begin{figure} 
\includegraphics[width=0.7\hsize,angle=-90]{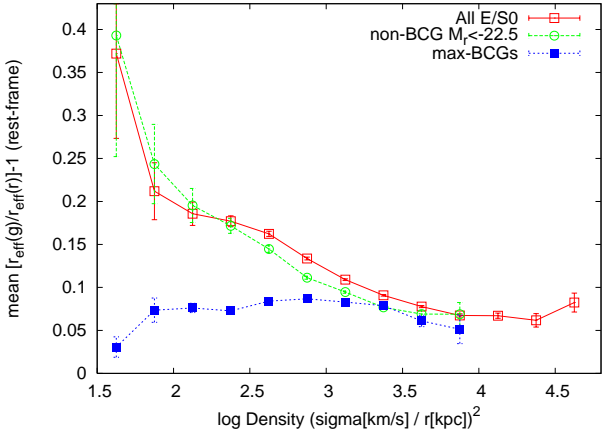} 
\caption{The mean ${{{\rm r}_{eff}(g)}\over{{\rm r}_{eff}(r)}}-1$ (corrected to rest-frame), for the full E/S0 sample, the $M_r<-22.5$ non-BCGs, and the max-BCGs, as a function of $\sigma^2/\rm r_{exp}^2$, a measure of mass density, expressed in a log scale for units of km $\rm s^{-1}$ and kpc.} 
 \end{figure}
 The full E/S0 sample and the $M_r<-22.5$ non-BCGs show a similar decrease in colour gradient with increasing mass density, whereas for BCGs ${{{\rm r}_{eff}(g)}\over{{\rm r}_{eff}(r)}}-1\simeq 0.08$ over the whole density range.
 
Figure 15 shows colour gradient against density for E/S0s divided by absolute magnitude. The negative correlation with density is seen over a very wide luminosity range, and strongest at $-21>M_r>-23$. 

To briefly summarize at this point, we find colour gradients in E/S0s to
(i) be greatest at intermediate luminosities and (ii) at a fixed luminosity to be anticorrelated with velocity dispersion and mass density, and that (iii) colour gradients tend to be flatter in BCGs and to show much less
dependence on galaxy properties. We now investigate the effects of age. 
\begin{figure} 
\includegraphics[width=0.7\hsize,angle=-90]{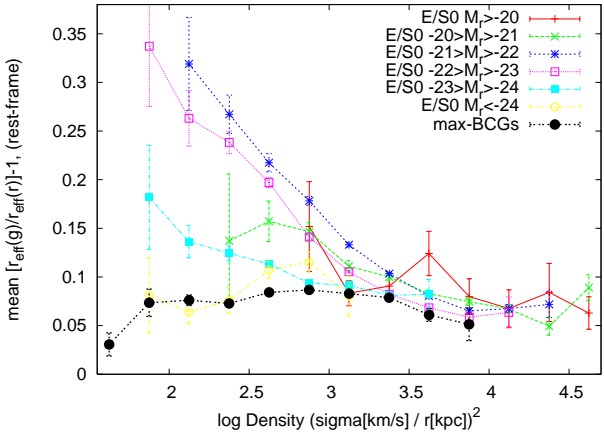} 
\caption{The mean ${{{\rm r}_{eff}(g)}\over{{\rm r}_{eff}(r)}}-1$  for the full E/S0 sample divided by $M_r$, and for the max-BCGs, as a function of $\sigma^2/\rm r_{exp}^2$, a measure of mass density.} 
 \end{figure}

\section{Age Effects on Colour Gradient and Colour}
For most galaxies in the E/S0 sample (64774/70378) and the $M_r<-22.5$ non-BCG sample (16337/18225), we have mean stellar age (`age50') estimated from the spectral index analysis of Gallazzi et al. (2005) performed on the 
$3.7\times 10^5$ galaxies of the DR4. This catalog includes only part of the max-BCG list. By position-matching the two lists we find there are stellar age estimates for almost half (2274/4919) the max-BCG sample, sufficient to investigate age-dependent effects. These stellar ages will generally be less than the lookback time to galaxy formation, and as they are luminosity-weighted, the age estimate of a galaxy may be greatly reduced if it has experienced recent episodes of star-formation.

The stellar mean age (and scatter) are 6.92 (1.74) Gyr for all E/S0s,
7.04 (1.69) Gyr for the non-BCGs, 7.54 (1.39) Gyr for the max-BCGs and 8.12 (1.33) Gyr for the C4 BCGs. The greater mean age of the C4 BCGs is due to these being a lower redshift sample than the others. Figure 16a shows the age distributions for the 3 deep samples; the modal ages are similar but there are fewer younger ($<6$ Gyr) and more older ages (9--10 Gyr) ages amongst the BCGs. 

We add the mean stellar ages to the lookback times at the redshifts of observation, giving lookback times to the mean stellar formation time, and then convert these back to `formation' redshifts (not the redshift of galaxy formation but the mean redshift of luminosity-weighted star formation, which will be lower) for the galaxies.
Figure 16b shows the distribution of the mean redshift of star formation; there is more of a difference here between the all-E/S0 sample and the BCGs and other high luminosity galaxies, as the more luminous galaxies tend to be observed at higher redshifts as well as being older. The mean star-formation redshift is 1.32 for all E/S0, 1.56 for the non-BCGs, 1.53 Gyr for the C4 BCGs and 1.82 for the max-BCGs; mean lookback times are 8.47, 9.17, 9.22 and 9.85 Gyr.

Figure 17 shows mean age against $M_r$; the BCGs are not only older than other E/S0s but show less of an increase in age with luminosity.
\begin{figure} 
\includegraphics[width=0.7\hsize,angle=-90]{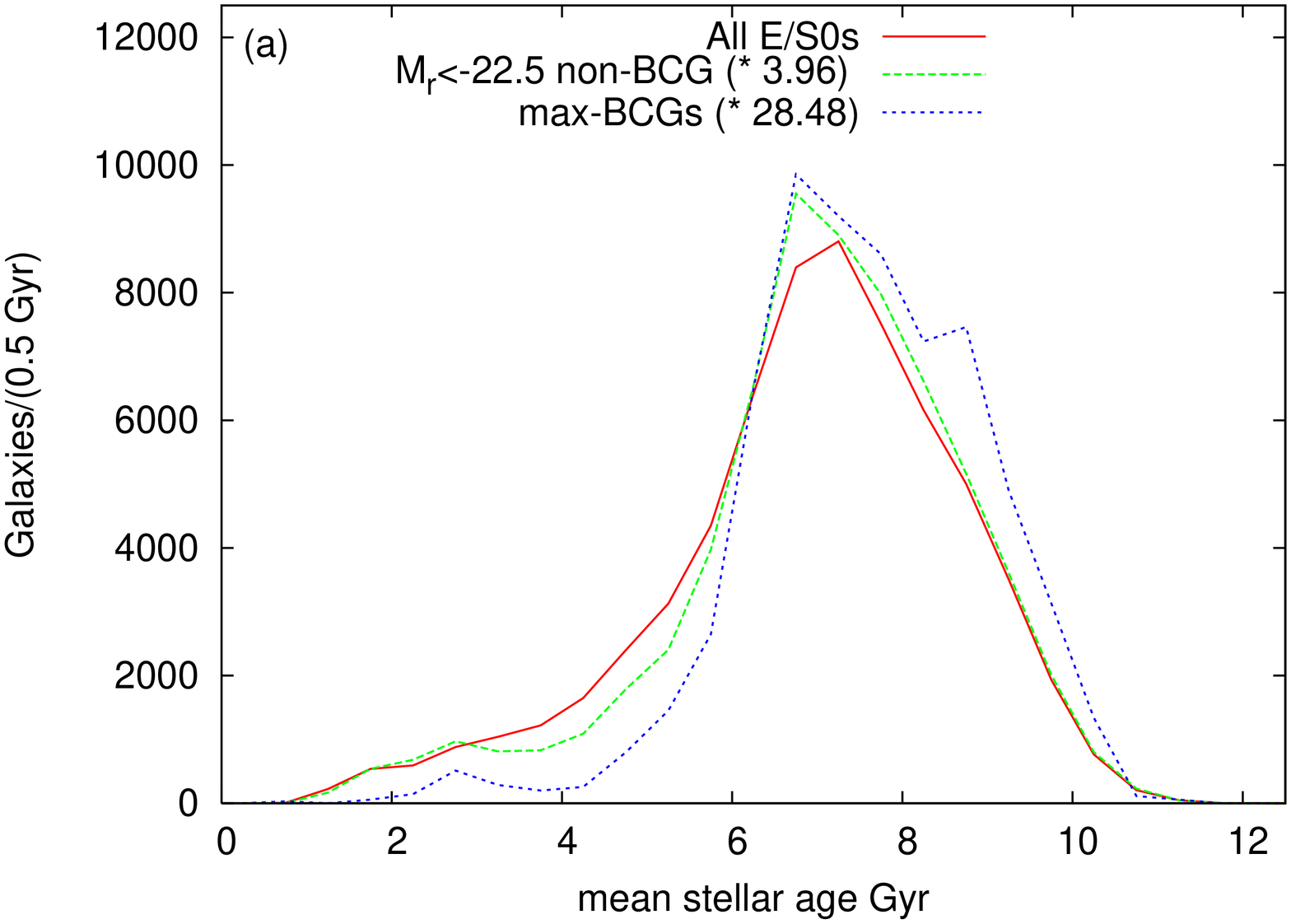}
 \includegraphics[width=0.7\hsize,angle=-90]{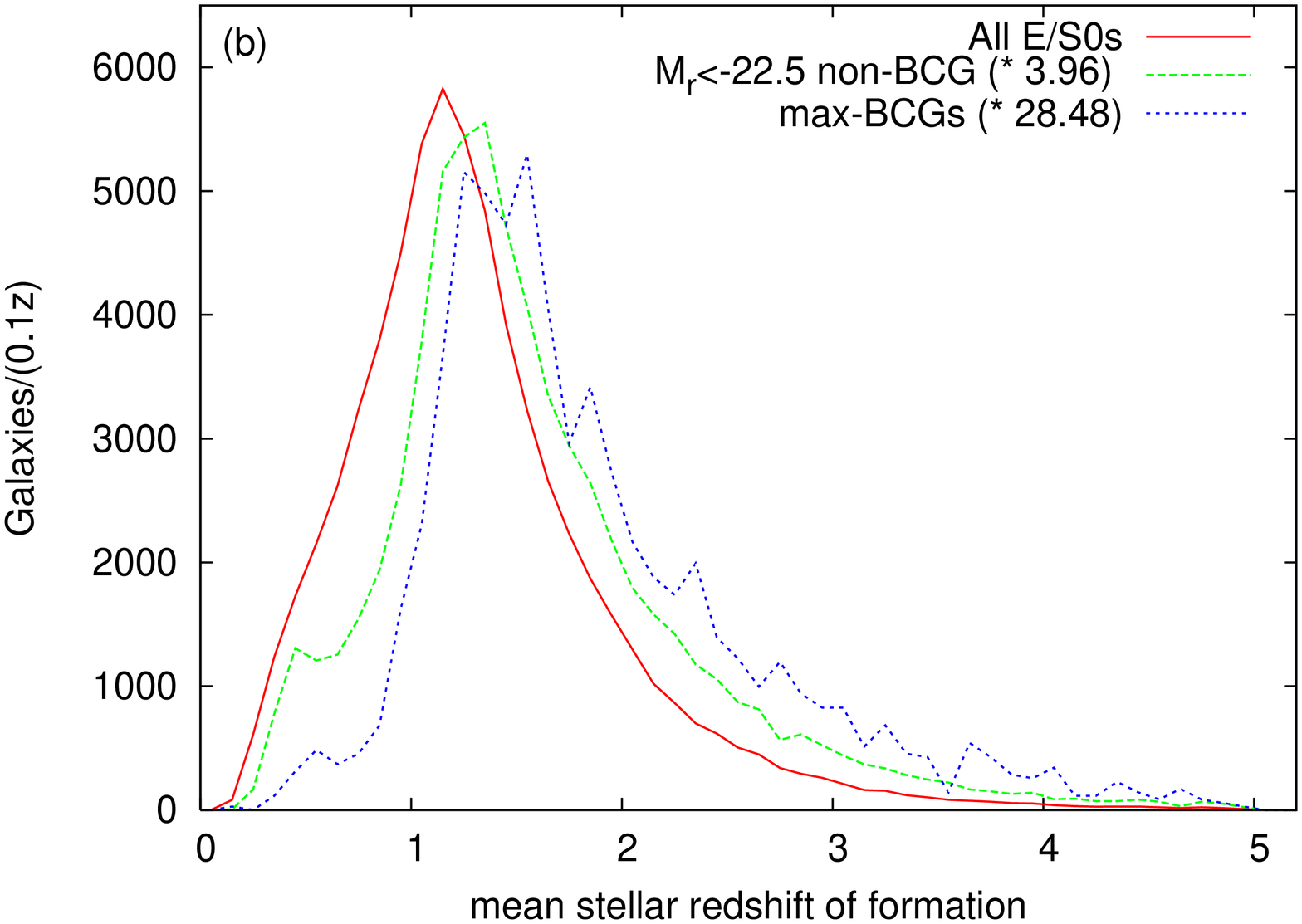}
\caption{Distribution of the (a) mean stellar age (b) mean redshift of star formation, for the full E/S0 sample, the $M_r<-22.5$ non-BCGs, and the max-BCGs, normalized to the same area under the curves.} 
 \end{figure}
\begin{figure} 
\includegraphics[width=0.7\hsize,angle=-90]{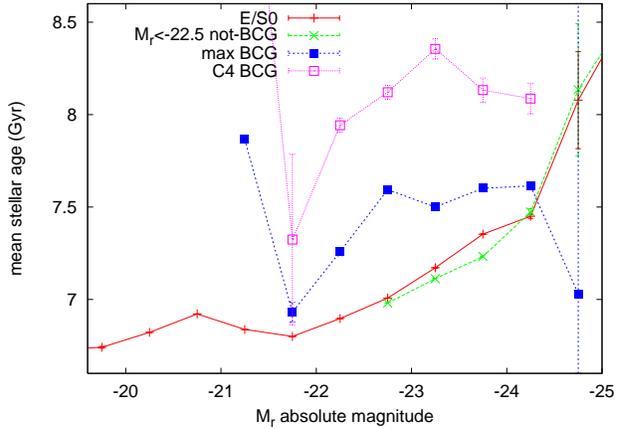} 
\caption{The mean stellar age against absolute magnitude $M_r$ for the full E/S0 sample, the $M_r<-22.5$ non-BCGs, and the max-BCGs, and the less deep C4 BCG sample.} 
 \end{figure}

We investigate whether the redder  colours and weaker colour gradients of BCGs are related to their greater ages, and in addition, whether the trends in (non-BCG) colour gradients with galaxy properties are caused by age effects.  A higher luminosity or lower $\sigma$ relative to the mean $\langle\sigma|M_r\rangle$ relation, found (in Paper I and here) to be associated with a stronger colour gradient, is also correlated with  a younger stellar age (Forbes and Ponman 1999; Bernardi et al. 2005; Gallazzi et al. 2006). 
\begin{figure} 
\includegraphics[width=0.7\hsize,angle=-90]{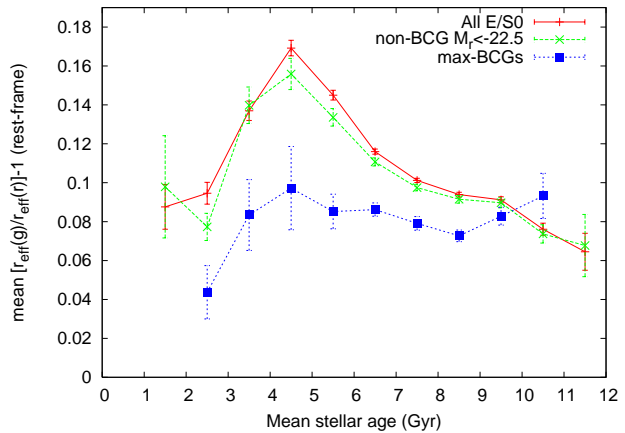} 
\caption{The mean colour gradient ${{{\rm r}_{eff}(g)}\over{{\rm r}_{eff}(r)}}-1$ (corrected to rest-frame) for the full E/S0 sample, the $M_r<-22.5$ non-BCGs, and the max-BCGs, as a function of mean stellar age ('age50') in Gyr,  for all galaxies for which we have age estimates from the methods of Gallazzi et al. (2005, 2006).} 
 \end{figure}
\begin{figure} 
\includegraphics[width=0.7\hsize,angle=-90]{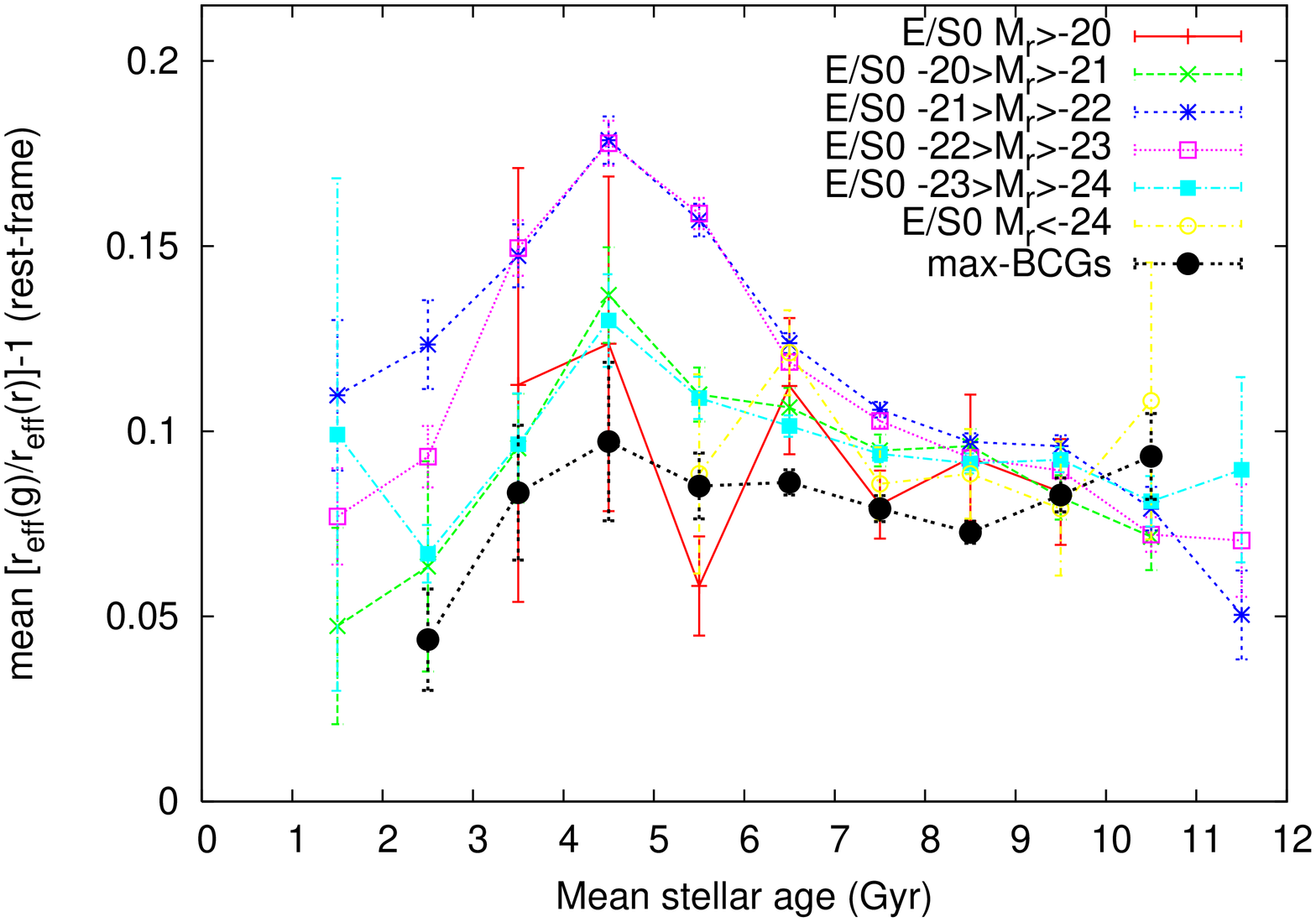} 
\caption{The mean colour gradient ${{{\rm r}_{eff}(g)}\over{{\rm r}_{eff}(r)}}-1$ (corrected to rest-frame) for the full E/S0 sample, as a function of mean stellar age ('age50') in Gyr, with galaxies divided by absolute magnitude $M_r$.} 
 \end{figure}

On Figure 18, for non-BCG E/S0s the colour gradients are maximum at stellar ages 4--5 Gyr and decrease by almost a factor of two to the highest ages. On Figure 19 it can be seen that this peak in gradient vs. age is most pronounced for galaxies of intermediate luminosities of $-21>M_r>-23$, weaker at both $-20>M_r>-21$ and $-23>M_r>-24$, and apparently absent at $M_r>-20$ and $M_r<-24$, following a rather symmetric luminosity trend. The BCG colour gradients show no age dependence (at $>4$ Gyr), and again the BCGs have lower colour gradients than the non-BCG E/S0 sample, at all but the highest ages where the two converge. 

A potential problem with this analysis is that the stellar ages are `central' ages estimated from the spectra in fixed apertures, and if colour gradients were typically driven by age gradients this would bias the age-gradient relation. We are not able to investigate age gradients with the single aperture of SDSS spectroscopy, but Tamura and Ohta (2004) and  La Barbera and Carvalho (2009) find that age gradients in early-types are typically very small or consistent with zero and the negative colour gradient is essentially a metallicity gradient (positive colour gradients can be stellar age related, as discussed below, but these occur in $<8\%$ of these samples).

At ages of $\leq 3$ Gyr there is sharp drop in mean colour gradient in the youngward direction, seemingly for all luminosities of E/S0 and even the BCGs (although very few - $1.1\%$ - of this BCG sample are this young). This can be explained by the younger stellar populations within these galaxies being more centrally concentrated, giving positive age gradients which counteract negative metallicity gradients to flatten the colour gradients. Some post-starburst galaxies have significantly opposite gradients i.e. blue cores (Yamauchi and Goto 2005) from central populations of very young stars (Suh et al. 2010).

 We examined the images and spectra of some of the galaxies with very inverted colour gradients. Although some of the most extreme examples were galaxies confused with stellar images, out of 40 galaxies with gradient $<-0.1$ and age $<3$ Gyr, most looked like normal ellipticals and many had strong Balmer absorption lines indicative of post-starburst galaxies. The best 7 examples had equivalent widths of 5--$8\rm \AA$ for both $\rm H\beta$ and $\rm H\delta$ absorption (these are J034453.17-065442.9, J150020.11+040000.2, J211348.16-063059.6, J210258.87+103300.6, J225226.07+150303.1, J130029.44+545503.9 and J155435.53+291319.9); there were a further 6 with $\rm H\beta$ $\rm EW=4.5$--$5.0\rm \AA$, and two more with less absorption in $\rm H\beta$  but $\rm H\delta$ $\rm EW>5 \AA$.

Figures 18 and 19 show that differences in the colour gradient properties of BCGs and non-BCGs are still apparent when they are age-matched, and so are not simply caused by the difference in their mean age. In the same way we can investigate whether this is the case for their colours.
Figure 20a shows mean model $g-r$ against stellar age; the youngest galaxies are much bluer, but the age dependence flattens at $>6$ Gyr, while the BCGs are redder than age-matched non-BCGs. The $M_r<-22.5$ non-BCGs are also redder than the full E/S0 sample, due to the higher metallicity of the more luminous galaxies (hence these are the more appropriate sample to compare colours with the BCGs), but the BCGs are $\sim 0.01$ mag redder still. Figure 20b shows how colour depends strongly on luminosity even when ages are matched (hence the CMR slope is not just age driven).
\begin{figure} 
\includegraphics[width=0.65\hsize,angle=-90]{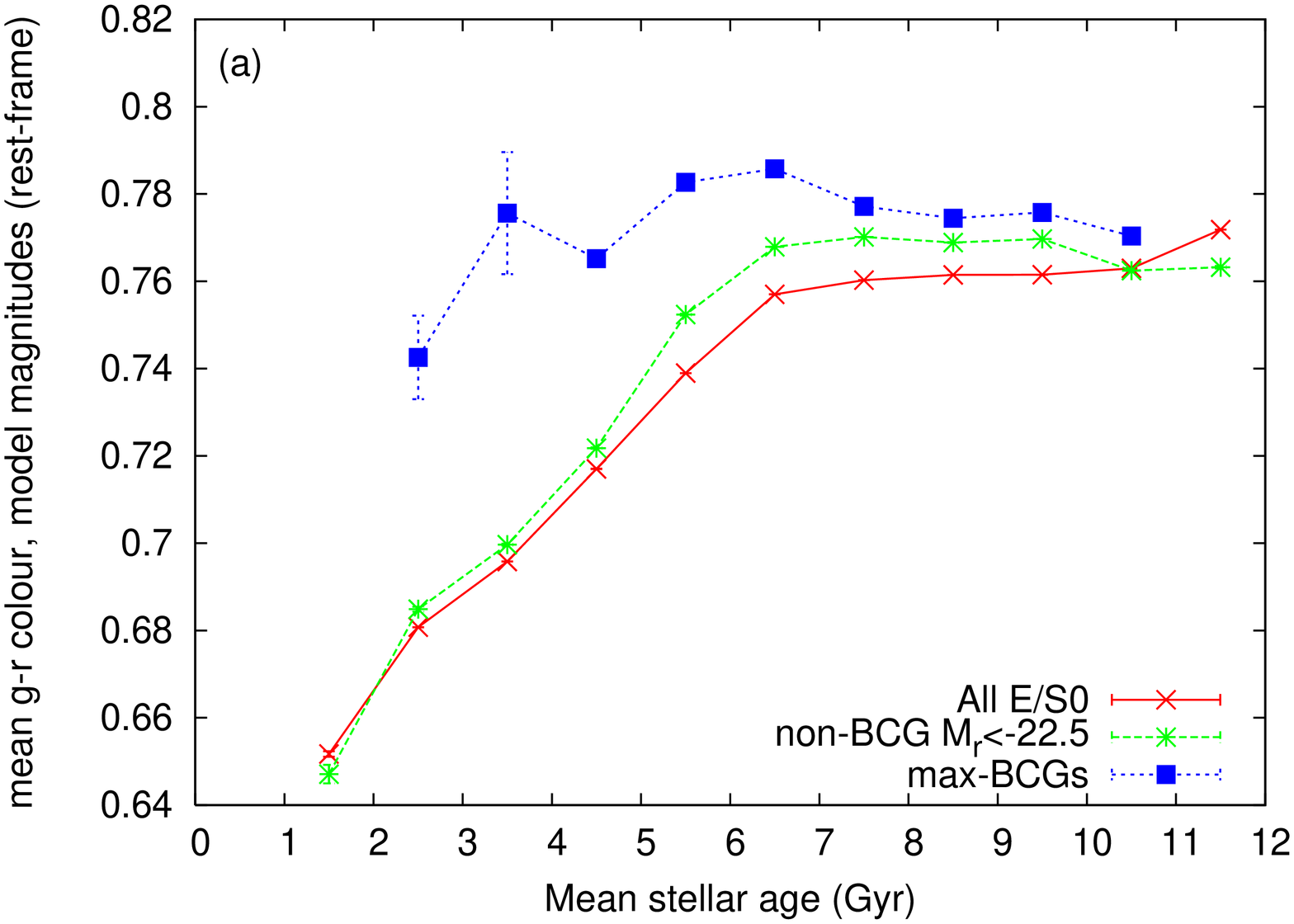}
 \includegraphics[width=0.65\hsize,angle=-90]{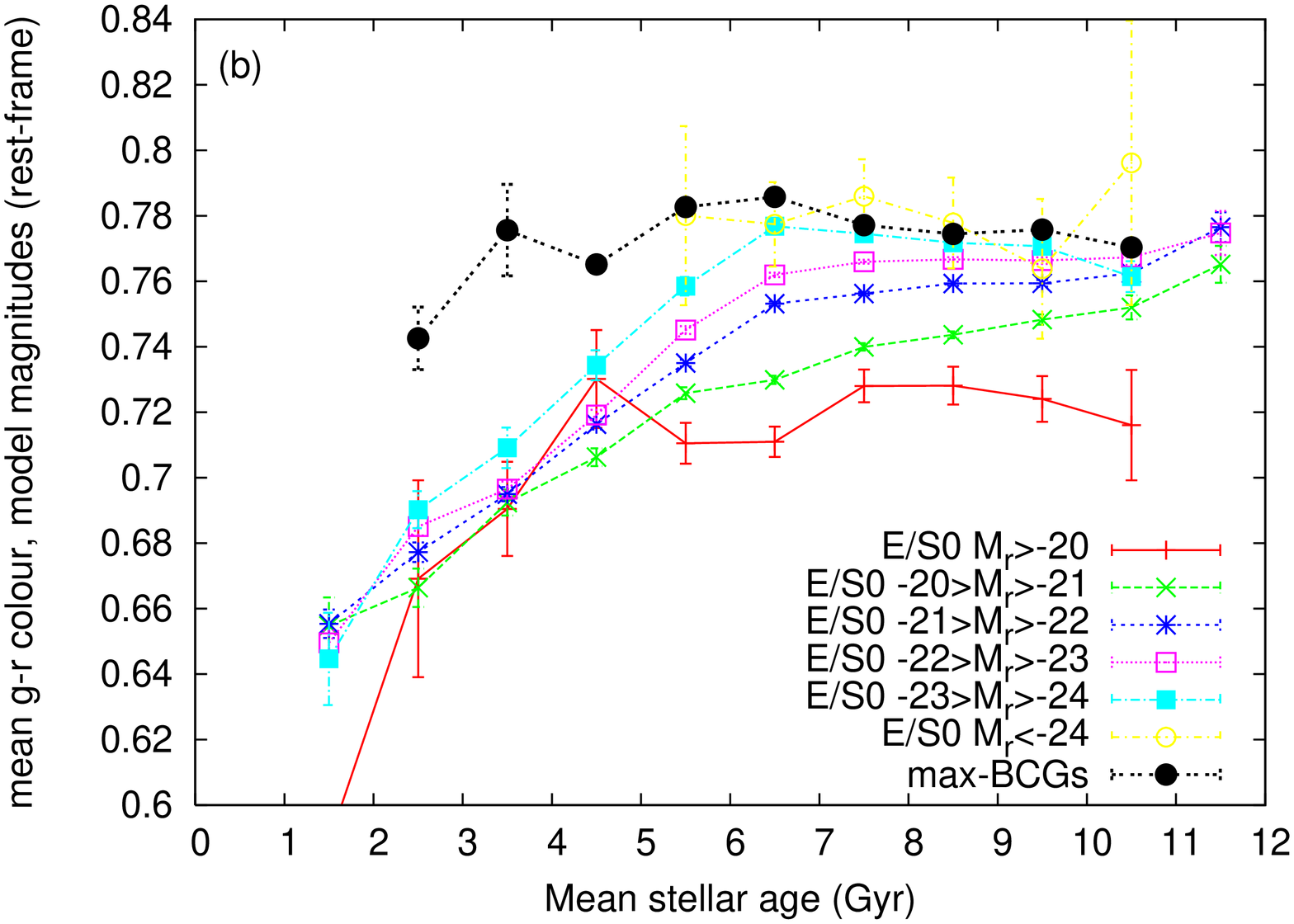}
\includegraphics[width=0.65\hsize,angle=-90]{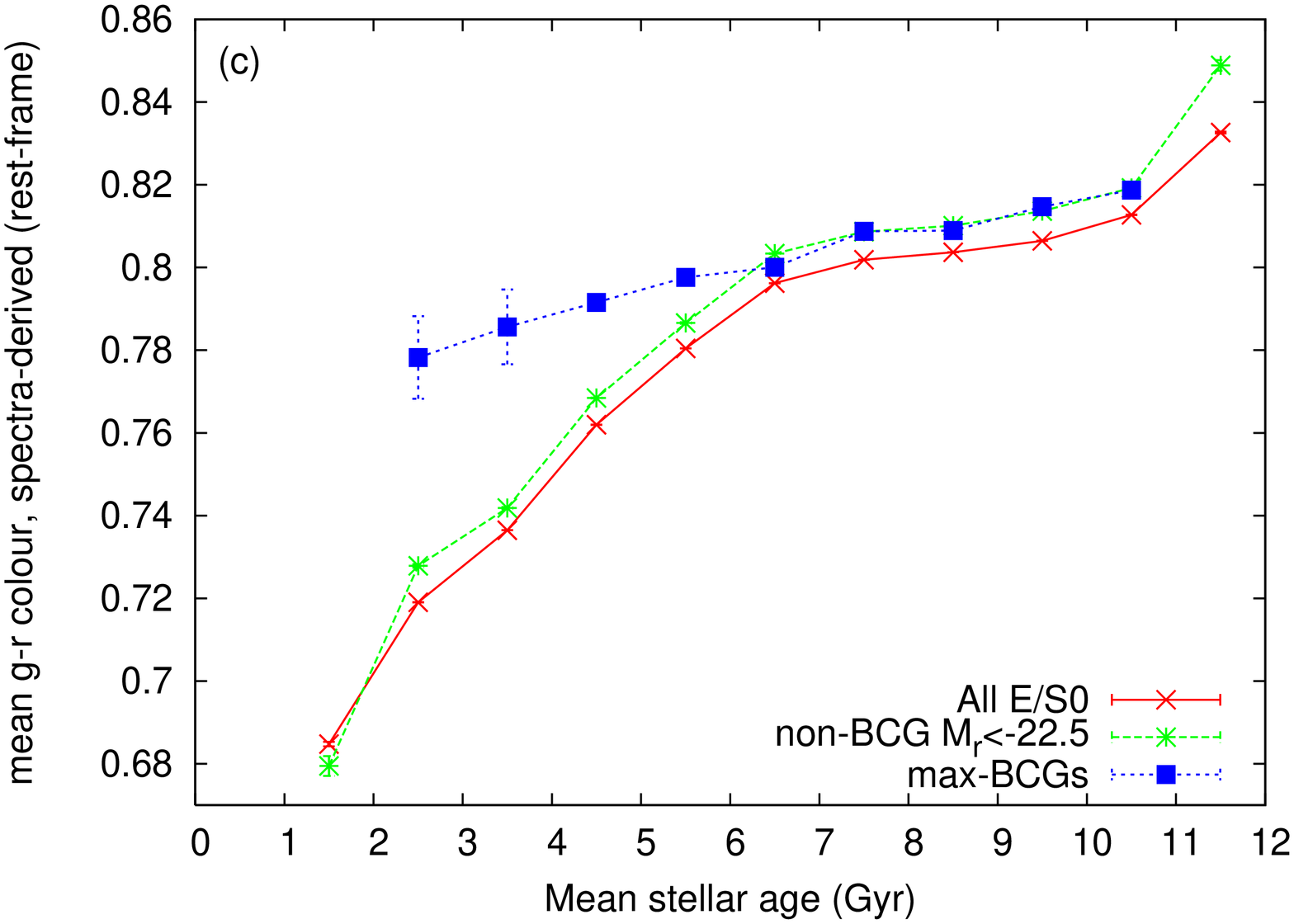}
 \includegraphics[width=0.65\hsize,angle=-90]{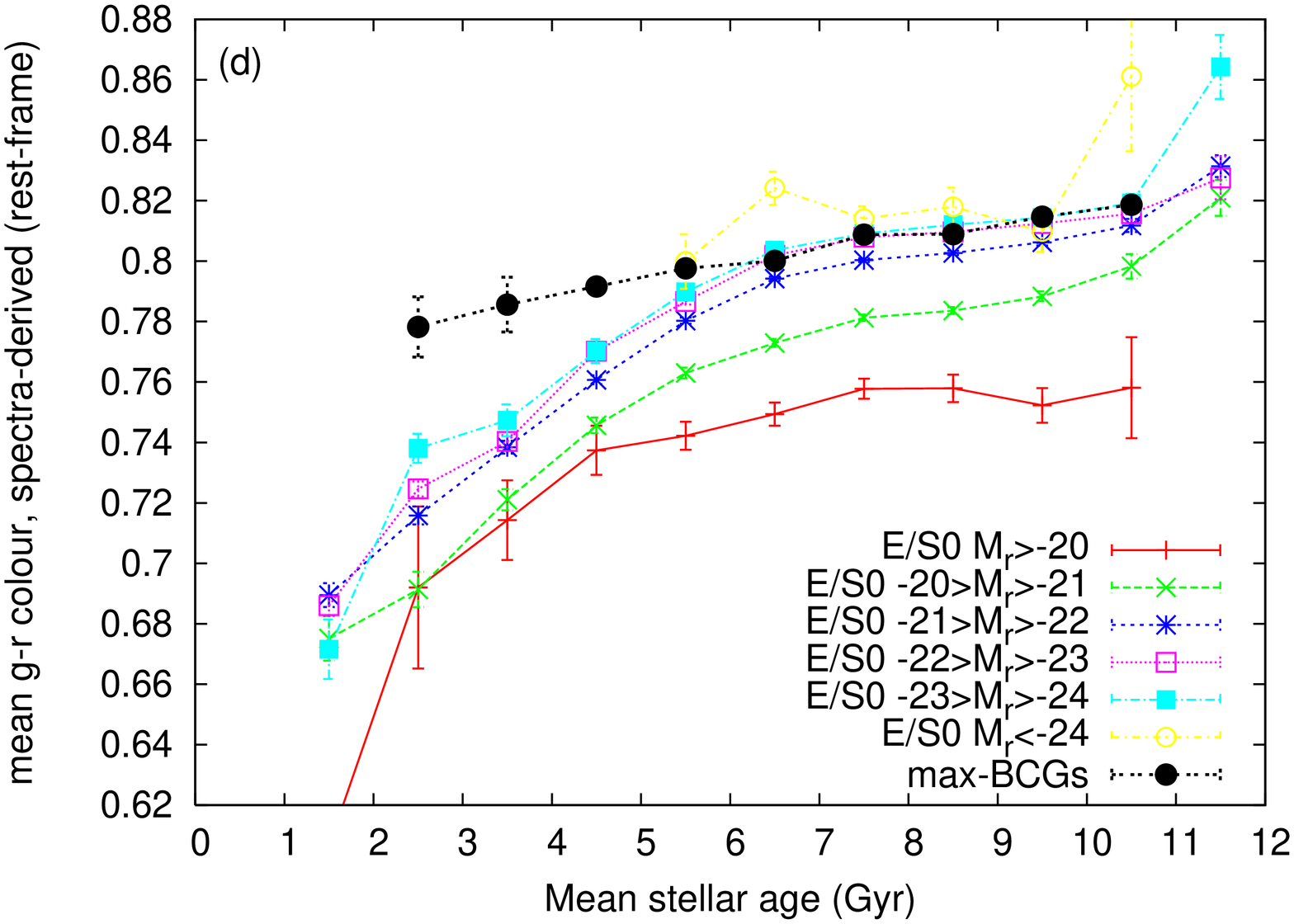}
\caption{The mean rest-frame $g-r$ from the k-corrected model magnitudes as a function of mean stellar age (age50) from Gallazzi et al. (2005), (a) for the full E/S0 sample, the $M_r<-22.5$ non-BCGs, and the max-BCGs, (b) for E/S0s divided by luminosity; (c,d) as (a,b) using spectra-derived magnitudes.} 
 \end{figure}

The spectra-derived colours (Figure 20c) follow a more linear relation to stellar age with $\Delta(g-r)\sim 0.5\Delta({\rm age})$, and, notably, do not appear to differ between BCGs and age-matched, similarly luminous non-BCGs (except for the few very young BCGs, which are redder). The difference between the model and spectra-derived (aperture) colours is caused by colour gradients.

We found BCGs to be at least $\Delta(g-r)\simeq 0.01$ mag redder than other E/S0s in the model-magnitude CMR and 
C$\sigma$R (Figures 1, 2, 3 and 4) with a smaller ($<0.005$ mag) or no significant difference in the spectra-derived colours. This can now be explained as a combination of 
(i) the 0.5 Gyr greater mean age of BCGs compared with luminosity-matched non-BCGs, corresponding to $\Delta(g-r)\simeq 0.0025$ in the aperture colour, and
(ii) the flatter colour gradients of BCGs, meaning that even if their spectra-derived (central) colours match those of non-BCGs, their model-magnitude colours will be redder. This could account for the remaining $\sim 0.0075$ magnitudes.

To further investigate the relationship between age and colour gradient,  we show mean ${{{\rm r}_{eff}(g)}\over{{\rm r}_{eff}(r)}}-1$ against radius (Figure 21) $10~{\rm log}~\sigma+M_r$ (Figure 22) and density (Figure 23) for the full E/S0 sample divided by mean stellar `formation' redshift. BCGs are included to see if their different properties with respect to other E/S0s resemble the effects of increased age. 
We also (Figure 24) plot colour gradient against dynamic mass, estimated as $5.8{{\rm r_{eff}\sigma^2}\over{G}}$; for units of km$\rm s^{-1}$ and kpc, 
${\rm log}~(M_{dyn}/M_{\odot})=\rm log~r_{eff} + 2 {\rm log}\sigma +6.13$.

Colour gradients show a broad peak at $M_{dyn}\simeq 10^{11.4}M_{\odot}$, which is more pronounced for galaxies with younger stellar ages.
\begin{figure} 
\includegraphics[width=0.7\hsize,angle=-90]{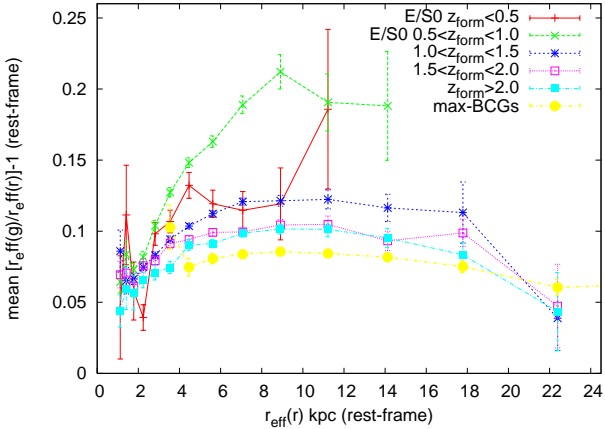} 
\caption{The mean colour gradient ${{{\rm r}_{eff}(g)}\over{{\rm r}_{eff}(r)}}-1$ (corrected to rest-frame) for the full E/S0 sample divided by mean stellar age (converted to a mean stellar formation redshift), and for the max-BCGs, as a function of radius.} 
 \end{figure}
\begin{figure} 
\includegraphics[width=0.7\hsize,angle=-90]{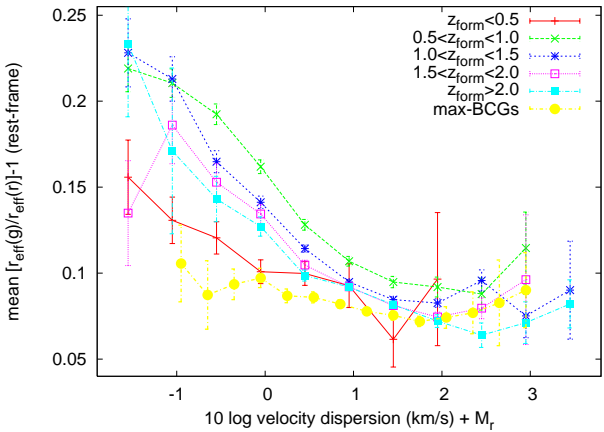} 
\caption{As Figure 21 as a function of the velocity dispersion to luminosity ratio.} 
 \end{figure}
\begin{figure} 
\includegraphics[width=0.7\hsize,angle=-90]{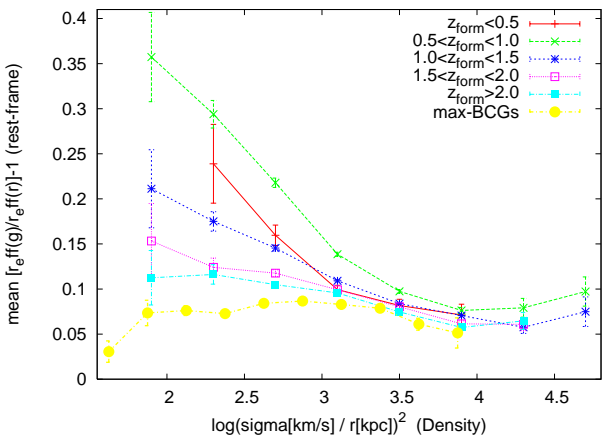} 
\caption{As Figure 21 as a function of mass density.} 
 \end{figure}
\begin{figure} 
\includegraphics[width=0.7\hsize,angle=-90]{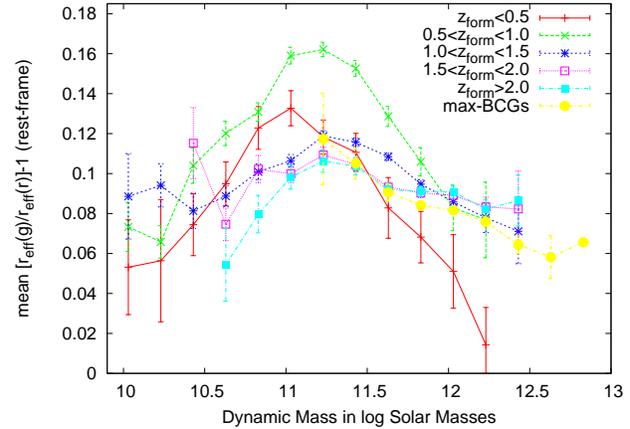} 
\caption{As Figure 21 as a function of dynamic mass ($5.8r_{eff}\sigma^2/G$).} 
 \end{figure}
 Colour gradients in the galaxies with the youngest stellar populations, $z_{form}<0.5$, show correlations with galaxy properties, similar to the $0.5<z_{form}<1.0$ galaxies, but offset downwards, as might result from the presence of an admixture of post-starbursts with negative gradients. 
Colour gradients in E/S0s depend significantly on radius, density, mass etc. {\it within} each $z_{form}$ interval. However, these correlations, especially those with radius and density, become less strong in $z_{form}>1.5$ galaxies. We discuss this further in the next Section.
\vfill
\eject
 \section{Summary and Discussion}
(i) The colour-magnitude relation (CMR) and colour-velocity dispersion relation (C$\sigma$R) of BCGs are significantly flatter in slope(${{d(g-r)}\over{d M_r}}$ and ${{d(g-r)}\over{d{\rm log}~\sigma}}$) than the respective relations for non-BCG E/S0 galaxies of comparable (high) luminosity. The difference in slope is about a factor of two, whether we define the $g-r$ colour using k-corrected model magnitudes or rest-frame magnitudes derived from integrating the SDSS spectra. BCGs are also, on average, $\sim 0.01$ mag redder in model magnitudes (k-corrected to rest-frame) than E/S0s of the same $M_r$ or $\sigma$, but with less or no offset in spectra-derived colours. 

 The hierarchical merging model of Lucia and Blaizot (2007) predicts a flat CMR for BCGs as a result of late assembly from a large number of red progenitors which formed their stars very much earlier ($z\sim 5$). This model included a very early truncation of star-formation. Skelton, Bell and Somerville (2009) present a simplified model in which dry mergers of already merged E/S0s on a `creation red sequence'  mildly flatten the CMR slope at higher luminosities. There may indeed be some flattening in the bright end of the CMR for all E/S0s (see Paper I), but for BCGs we find much more, although the BCG CMR is not entirely flat. This suggests the BCGs or their progenitors evolve by a less extreme scenario than the first of these models but perhaps with a higher dry merger rate compared to the E/S0s in the second model. In addition to being flattened, the BCG CMR is offset redwards.
\medskip

(ii) As a simple quantifier of radial colour gradient we use  ${{{\rm r}_{eff}(g)}\over{{\rm r}_{eff}(r)}}-1$, with the 
effective radii ${\rm r}_{eff}(g)$ and ${\rm r}_{eff}(r)$, corrected to the rest-frame by linearly interpolating between the radii fitted in the bands $g$, $r$, $i$ and $z$. We show that at least for galaxies with angular $r_{eff}$ radii above 1.8 arcsec, this ratio is well and linearly correlated to ${{d(g-r)}\over {d(\rm log~r)}}$ across the observed range of colour gradients and galaxy luminosities, and estimate 
 ${{d(g-r)}\over {d(\rm log~r)}}\simeq -0.87 ({{{\rm r}_{eff}(g)}\over{{\rm r}_{eff}(r)}}-1)$. The large scatter in this correlation, which we initially estimated as 0.122, is reduced to a much more acceptable 0.056 if only galaxies with $r_{eff}$ larger than 1.8 arcsec are included.

After excluding the $r_{eff}<1.8$ arcsec galaxies and a few others which we believed to be spiral types, we found for the E/S0s a mean colour gradient, using our measure, of $0.11115\pm 0.00061$ with a comparable galaxy-to galaxy scatter 0.1172. For the $M_r<-22.5$ non-BCG the mean is similar at $0.10550\pm 0.00098$ (scatter 0.1141) and for max-BCGs the mean  is $0.08142\pm 0.00114$ (scatter 0.0764). Therefore we find at high significance that BCGs have a flatter (by an average of $23\%$) radial colour gradient than other high luminosity spheroidals. 

This again could be due to intensified merging in the cluster-core environment. La Barbera et al. (2005), and similarly Ko and Im (2005), reported mean colour gradients in E/S0s of all luminosities were weaker by almost a factor two in the richest clusters ($N_{gal}>50$) compared to the field, and suggested this was because formation histories in denser cluster environments included more elliptical-elliptical mergers (e.g. Tran et al. 2005), which reduce colour/metallicity gradients (Kobayashi 2004). 

The simulations of Boylan-Kolchin, Ma and Quataert (2006) account for the steeper $R_e$--$L$ of BCGs, i.e. their `abnormally' large sizes (e.g. Bernardi 2009), as the result of repeated `dry' (dissipationless, non-star-forming) mergers, and especially of the anisotropic accretion of smaller spheroidals in near-radial, low angular momentum  orbits, which might occur preferentially in a cluster core. A few BCGs have been observed undergoing such mergers (Liu et al. 2008; Tran et al. 2008).
Di Matteo et al. (2009) predict from simulations that the metallicity gradient in a  dissipationless (`dry') merger remnant will consistently be $\simeq 0.6$ the mean of the progenitor galaxies' gradients (but not reduced to zero).
\medskip

(iii) In non-BCG E/S0s, we examine the trends in mean colour gradient as a function of other galaxy properties. We find a dependence of colour gradient on absolute magnitude $M_r$, with a broad peak in colour gradient at $M_r\simeq - 22$ and a decrease to lower and higher luminosities. 
The luminosity dependence is relatively mild, less than a factor 1.5, which may explain why it was not seen by La Barbera et al. (2005). However, Spolaor et al. (2009) do find a luminosity dependence similar to ours.
Colour gradients tend to decrease with increasing velocity dispersion $\sigma$, by almost 1/2  between 150 and 300 km $\rm s^{-1}$, confirming the suggestion in our Paper I that the trends with $M_r$ and $\sigma$ are different.
\medskip

(iv) Colour gradients in E/S0s show strong correlations with some other galaxy properties. Colour gradients increase, by about a factor of two, with from small to larger effective radius, up to a maximum at 8-12 kpc (depending on luminosity). A positive correlation with radius was prevously reported by Tamura and Ohta (2003) for a small sample of E/S0s in Abell 2199. At even larger radii we find the mean colour gradients decrease, but this is actually because of a reduced gradient in the most luminous ($M_r<-23$) galaxies. If E/S0s are divided by luminosity, within  low/moderate luminosity intervals there is simply a steep increase in colour gradient with $\rm r_{eff}$.

We find colour gradients to be negatively correlated with
$\rm 10~log~\sigma+M_r$ (the $\sigma$ residual relative to $\langle \sigma|M_r\rangle$)  and mass density ($\sigma^2/\rm r_{eff}^2$).  Of course, these quantities are related -- a high density implies a high $\sigma$ at a given dynamic mass (rather than luminosity). These negative correlations do not seem to be caused by a selection effect (from the flux limit of the sample), as they are seen within a wide range of narrow luminosity intervals, being strongest for E/S0s of moderate/high luminosity $-21<M_r<-24$. 
\medskip 
 
(v) We also examine the trend in colour gradient with the `mean stellar ages' estimated (Gallazzi et al. 2005, 2006) for most of these galaxies using 5 indices from the SDSS spectra. In non-BCG E/S0s the colour gradients are strongest for relatively young ages of 4--5 Gyr
(which may signify the presence of younger stellar populations or more extended star-formation, rather than the galaxies being younger) and decrease by up to a factor of two to the oldest ages of 11 Gyr.
This may be primarily because older stellar populations have experienced more mergers and interactions, especially at high redshifts. Early-type galaxies which ceased forming stars and joined the red sequence more recently (e.g. post-mergers of spirals at $z<1$) would tend to have stronger colour gradients, if it is dry mergers (occurring after the end of star-formation) that flatten the gradients.

However, there is a drop in mean colour gradient for the galaxies with mean stellar ages less than 3--4 Gyr (even in BCGs). This is likely due to the inclusion of post-starburst galaxies, which can have blue cores giving inverted (positive) colour gradients (Friaca and Terlevich 1999; Yamauchi and Goto 2005) together with young age indicators like strong Balmer absorption lines. To confirm this we examined the spectra of 40 galaxies with age $<3$ Gyr and inverted colour gradients ($<-0.1$) and found many with enhanced Balmer absorption including 7 with equivalent widths $>5\rm \AA$  in both $\rm H\beta$ and $\rm H\delta$.
 Suh et al. (2010) found that blue-cored early-types in general had at least mildly enhanced $\rm H\beta$ absorption, and showed that the Balmer enhancement and the positive colour gradients could both be accounted for by centrally concentrated, very young stellar populations (age $\leq \rm 0.5 Gyr$), comprising only $\sim0.5$--$2\%$  of the stellar mass.
\medskip

(vi) Colour gradients in BCGs are consistently lower than in non-BCG spheroidals of the same luminosity, velocity dispersion or radius.
 Some classes of non-BCG  ellipticals, those with (i) the highest luminosities and largest radii, (ii) the highest $\sigma$ relative to $M_r$, or (iii) the highest densities, also have mean colour gradients of ${{{\rm r}_{eff}(g)}\over{{\rm r}_{eff}(r)}}\simeq 0.08$. But within the BCG class mean ${{{\rm r}_{eff}(g)}\over {{\rm r}_{eff}(r)}}-1$ remains at $\sim 0.08$, almost regardless of the other galaxy properties. The mean colour gradient in BCGs decreases a little for the largest radii and with increasing dynamic mass, but is uncorrelated with mass density, the $10~{\rm log}~\sigma+M_r$ residual, or stellar age (except for the drop at $<3$ Gyr). 

The process of colour gradient flattening through dry mergers appears to have operated most strongly on (all) BCGs, reducing the mean colour gradients to $\sim 0.08$, but never much lower than this, whatever the initial value. In the Di Matteo et al. (2009) model of dry mergers, a galaxy with a strong colour gradient will probably merge with one of lower gradient and experience a significant flattening, but a galaxy with already a very flat colour gradient would probably merge with one with higher gradient and experience no further decrease or even a slight increase. With repeated mergers this would cause an exponential decrease in the mean gradient and also a narrowed distribution, which is what we seem to find for the merged classes of galaxies, with mean gradients of $\simeq 0.08$ consistently. We found the BCGs had ${1\over 3}$ less galaxy-to-galaxy variation in colour gradient than the E/S0s (this difference might be even greater after taking measurement errors into account).

In addition, the radial mergers thought to occur in BCGs would strongly  increase their radii (and reduce density) while simultanously flattening colour gradients, thus erasing the gradient-radius correlation (and gradient-density anticorrelation) seen in other spheroidals.
\medskip

(vii) We examine colour as a function of stellar age, comparing the BCGs and other E/S0s. The BCGs are on average 0.5 Gyr older at observation than the $M_r<-22.5$ non-BCGs, and their mean lookback time to formation is estimated as 0.68 Gyr greater. This could simply reflect the correlation of earlier formation (by as much as $\sim 1$ Gyr) with high-density cluster environment (e.g. Sheth et al. 2006; Rogers et al. 2010). The BCGs also have a flatter age-luminosity relation, as might be expected if BCGs (of all masses) form from dry major mergers of red early-type galaxies (with older ages associated with the cluster-core environment), which would increase luminosity without changing the mean stellar age. In the same way, their CMR is flattened. If, on the other hand, the BCGs had formed `monolithically', without major merging, they might show stronger trends in their stellar populations (age, composition) with luminosity/mass, extrapolating the trends seen in less massive E/S0s.

 In rest-frame model-magnitude $g-r$, BCGs are slightly redder than non-BCG E/S0s of the same age, whereas in spectra-derived $g-r$, they lie on the same (approximately linear) relation of colour to age.
 We conclude that the ($\simeq 0.01$ mag) redder model-magnitude $g-r$ colours of the BCGs in the CMR and C$\sigma$R can be explained simply by their greater mean ages combined with their flatter colour gradients, which would give them a slightly redder model-magnitude colour for a given central-aperture (e.g. spectra-derived) colour. 
\medskip

(viii) We re-examine the correlations of colour gradient with galaxy properties, dividing by age (star-formation redshift). The positive correlation of colour gradient with radius, and the anticorrelation with density and $10~{\rm log}~\sigma +M_r$ are strongest in younger galaxies but are seen within all age intervals. With increasing age, these correlations flatten somewhat while the mean colour gradients decrease, as might be expected for sequential dry mergers in the Di Matteo et al. (2009) model.

The decrease in colour gradient with the $\sigma$ residual $10~{\rm log}~\sigma+M_r$ is not solely due to the known correlation of this quantity with stellar age (Forbes and Ponman 1999; Bernardi et al. 2005; Gallazzi et al. 2006), as it is seen within each age interval. Rather, it seems the colour gradient must have independent anticorrelations with the $\sigma$ to luminosity  ratio and with stellar age. 

It seems the process of spheroid formation sets up a  metallicity/colour gradient, with a steepness positively correlated with the mass/luminosity of stars and also with a large radius and low mass density for the galaxy. 
Subsequent mergers flatten the stronger gradients while adding low-gradient galaxies at the high-luminosity end of the E/S0 luminosity function. As dry mergers tend to increase size this also weakens the gradient-density relation (Figure 23). A comparison with colour gradients in much higher redshift E/S0s would help to confirm this. The mean colour gradient in E/S0s decreases somewhat at $M_{dyn}>10^{11.5}M_{\odot}$ because most of these higher-mass galaxies are the products of elliptical-elliptical (dry) mergers (Tran et al. 2005; Naab, Khochfar and Burkert 2006). Yet some E/S0s remain today with colour gradients more than twice the observed average, like those in the `monolithic' models of Kobayashi et al. (2004).

 The position of BCGs on Figures 21 to 24 is that of galaxies which have experienced more age and/or merger-related colour gradient suppression than even the $z_{form}>2$ E/S0s. On Fig 18 the BCG colour gradient matches that of non-BCGs of mean stellar age $>10$ Gyr, although the BCG mean stellar age is only 7.5 Gyr. The colour gradients in BCGs could be effectively `super-aged' by several Gyr by their environment and their position in the path of infalling cluster spheroidals.

 \section*{Acknowledgements} 
The authors are grateful for support provided by NASA grant ADP/NNX09AD02G.

Funding for the Sloan Digital Sky Survey (SDSS) has been provided by the Alfred P. 
Sloan Foundation, the Participating Institutions, the National Aeronautics and Space Ad- 
ministration, the National Science Foundation, the U.S. Department of Energy, the Japanese 
Monbukagakusho, and the Max Planck Society. The SDSS Web site is http://www.sdss.org/. 
The SDSS is managed by the Astrophysical Research Consortium (ARC) for the Participat- 
ing Institutions. The Participating Institutions are The University of Chicago, Fermilab, the 
Institute for Advanced Study, the Japan Participation Group, The Johns Hopkins University, 
the Korean Scientist Group, Los Alamos National Laboratory, the Max-Planck-Institute for 
Astronomy (MPIA), the Max-Planck-Institute for Astrophysics (MPA), New Mexico State 
University, University of Pittsburgh, University of Portsmouth, Princeton University, the 
United States Naval Observatory, and the University of Washington.

\section*{References} 
\vskip0.15cm \noindent Adelman-McCarthy et al., 2006, ApJS, 162, 38.

\vskip0.15cm \noindent Adelman-McCarthy et al. 2008, ApJS, 175, 297.

\vskip0.15cm \noindent Bernardi M., 2009, MNRAS, 395, 1491.

\vskip0.15cm \noindent Bernardi M., et al. 2003, AJ, 125, 1882.

\vskip0.15cm \noindent Bernardi M., Sheth R.K., Nichol R.C., Schneider 
D.P., Brinkmann J., 2005, AJ, 129, 61.

\vskip0.15cm \noindent Bernardi M., Nichol R.C., Sheth R.K., Miller 
C.J., Brinkmann J., 2006, AJ, 131, 1288.

\vskip0.15cm \noindent Bernardi M., Hyde J.B., Sheth R.K., Miller C.J., 
Nichol R.C., 2007, AJ, 133, 1741.

\vskip0.15cm \noindent Boylan-Kolchin M., Ma Chung-Pei, Quataert E., 2006, MNRAS, 369, 1081.

\vskip0.15cm \noindent Bruzual A.G. and Charlot S., 2003, MNRAS, 344, 
1000.

\vskip0.15cm \noindent De Lucia G. and Blaizot J., 2007, MNRAS, 375, 2.

\vskip0.1cm \noindent Di Matteo P., Pipino A., Lehnert M.D., Combes F., Semelin B., 2009, A\&A, 499, 427.

\vskip0.15cm \noindent Gallazzi A., Charlot S., Brinchmann J., White 
S.D.M., Tremionti C.A., 2005, MNRAS, 362, 41.

\vskip0.15cm \noindent Gallazzi A., Charlot S., Brinchmann J., White 
S.D.M., 2006, MNRAS, 370, 1106.

\vskip0.15cm \noindent Forbes D.A., Ponman T.J., 1999, MNRAS, 309, 623.

\vskip0.15cm \noindent Friaca A.C.S., Terlevich R.J., 1999, MNRAS, 325, 335.

\vskip0.15cm \noindent Hogg D., Baldry I., Blanton M., Eisenstein D., 2002, astro-ph/0210394

\vskip0.15cm \noindent Hyde J. and Bernardi M., 2009, MNRAS, 394, 1978.
 
\vskip0.15cm \noindent Jimenez R., Mariangela B., Haiman Z., Panter B., 
Heavens A.F., 2007, ApJ, 669, 947.

\vskip0.15cm \noindent J{\/o}rgensen I., Franx M., Kjaergaard P., 1995, MNRAS, 276, 1341.

\vskip0.15cm \noindent Ko J. and Im M., 2005, Journal of the Korean
 Astronomical Society, 38, 149.

\vskip0.15cm \noindent Kobayashi C., 2004, MNRAS, 347, 740.

\vskip0.15cm \noindent Koester B.P. et al. 2007, ApJ, 660, 239.

\vskip0.15cm \noindent La Barbera F., Carvalho R.R., Gal R.R., Busarello G., Merluzzi P., Capaccioli M., Djorgovski S.G., 2005, ApJ, 626, L19.
 
\vskip0.15cm \noindent La Barbera F., Carvalho R.R., 2009, ApJ, 699, L76.

\vskip0.15cm \noindent Liu F.S, Xia X.Y, Shude M., Wu H., Deng Z.G., 
2008, MNRAS, 385, 23.

\vskip0.15cm MehlertD., Thomas D., Saglia R.P., Bender R., Wegner G.,
2003, A\&A, 407, 423.

\vskip0.15cm \noindent Miller C., et al. 2005, AJ, 130, 968.

\vskip0.15cm Michard R., 2005, A\&A, 441, 451. 

\vskip0.15cm \noindent Naab T., Khochfar S., Burkert A., 2006, ApJ, 636, 81.

\vskip0.15cm \noindent Roche N. Bernardi M., Hyde J. 2009, MNRAS, 398, 
1549 (Paper I).

\vskip0.15cm \noindent Rogers B., Ferreras I., Pasquali A., Bernardi M.,
 Lahav O., Kaviraj S., 2010, MNRAS, in press (astro-ph/1002.0835)

\vskip0.15cm \noindent Sheth R.K., Jimenez R., Panter B., Heavens A.F., 2006, ApJ, 650, 1, L25.

\vskip0.15cm \noindent Skelton R.E., Bell E., Somerville R.S., 2009,
ApJ, 699, L9.

\vskip0.15cm \noindent Spolaor M., Proctor R.N., Forbes D., Couch W.J., 2009, ApJ, 691, L138. 

\vskip0.15cm \noindent Suh H., Jeong H., Oh K., Yi S.K, Ferras I., 
Schawinski K., 2010, ApJS, in press. (astro-ph/1002.3424)

\vskip0.15cm \noindent Tamura N., Kobayashi C., Arimoto N., Kodama T., Kouji O., 2000, AJ, 119, 2134.

\vskip0.15cm \noindent Tamura N., Ohta K., 2003, AJ, 126, 596.

\vskip0.15cm \noindent Tamura N., Ohta K., 2004, MNRAS, 355, 617.

\vskip0.15cm \noindent Tran K.-V., van Dokkum P., Franx M., Illinhworth G., 
Kelson D., Natascha M., Schreiber F., 2005, ApJ, 627, L25.

\vskip0.15cm \noindent Tran K.-V., Moustakas J., Gonzalez A.H., Anthony H., Bai L., Zaritsky D., Kautsch S.J., 2008, ApJ, 683.

\vskip0.15cm \noindent van der Wel A., Holden B., Zirm A.W., Franx M.,
Rettura A., Illingworth G., Ford H., 2008, ApJ, 688, 48.

\vskip0.15cm \noindent Wu H., Shao Z., Xia X., Deng Z., 2005, ApJ, 622, 244.

\vskip0.15cm \noindent Yamauchi C., Goto T., 2005, MNRAS, 359, 1557.

\end{document}